\documentclass[prd,twocolumn,amsmath,amssymb,floatfix,superscriptaddress,notitlepage,nofootinbib,preprintnumbers]{revtex4-1}
\usepackage{amsfonts,amssymb,amsmath,graphicx,color,bm,enumitem}
\definecolor{ultramarine}{rgb}{0.07, 0.04, 0.56}
\definecolor{cadmiumgreen}{rgb}{0.0, 0.42, 0.24}
\definecolor{indigo(dye)}{rgb}{0.0, 0.25, 0.42}
\usepackage[linktocpage=true]{hyperref}
\hypersetup{
colorlinks=true,
citecolor=ultramarine,
linkcolor=cadmiumgreen,
urlcolor=indigo(dye),
}

\newcommand{\f}[2]{\frac{#1}{#2}}  
\newcommand{\mk}[1]{\left( #1 \right)}  
\newcommand{\kk}[1]{\left[ #1 \right]}  
\newcommand{\ck}[1]{\left\{ #1 \right\}}  
\newcommand{\be}{\begin{equation}}  
\newcommand{\ee}{\end{equation}}
\newcommand{\Mpl}{M_{\rm Pl}}
\newcommand{\E}{\mathcal{E}}
\renewcommand{\L}{\mathcal{L}}
\newcommand{\pa}{\partial}

\begin{document}

\preprint{YITP-18-27}

\title{  
General Relativity solutions in modified gravity
}

\author{Hayato Motohashi}
\affiliation{Center for Gravitational Physics, Yukawa Institute for Theoretical Physics, Kyoto University,\\ Kyoto 606-8502, Japan}

\author{Masato Minamitsuji}
\affiliation{Centro de Astrof\'{\i}sica e Gravita\c c\~ao  - CENTRA,
Departamento de F\'{\i}sica, Instituto Superior T\'ecnico - IST,
Universidade de Lisboa - UL, Av. Rovisco Pais 1, 1049-001 Lisboa, Portugal}

\begin{abstract}
Recent gravitational wave observations of binary black hole mergers and a binary neutron star merger by LIGO and Virgo Collaborations associated with its optical counterpart constrain deviation from General Relativity (GR) both on strong-field regime and cosmological scales with high accuracy, and further strong constraints are expected by near-future observations.  Thus, it is important to identify theories of modified gravity that intrinsically possess the same solutions as in GR among a huge number of theories. We clarify the three conditions for theories of modified gravity to allow GR solutions, i.e., solutions with 
the metric satisfying the Einstein equations in GR and the constant profile of the scalar fields. Our analysis is quite general, as it applies a wide class of single-/multi-field scalar-tensor theories of modified gravity in the presence of matter component, and any spacetime geometry including cosmological background as well as spacetime around black hole and neutron star, for the latter of which these conditions provide a necessary condition for no-hair theorem. The three conditions will be useful for further constraints on modified gravity theories as they classify general theories of modified gravity into three classes, each of which possesses i) unique GR solutions (i.e., no-hair cases), ii) only hairy solutions (except the cases that GR solutions are realized by cancellation between singular coupling functions in the Euler-Lagrange equations), and iii) both GR and hairy solutions, for the last of which one of the two solutions may be selected dynamically.
\end{abstract}

\maketitle  


\section{Introduction} \label{sec1}
Recent measurements of gravitational waves (GWs) 
from binary black hole (BH) mergers 
by LIGO and Virgo Collaborations
\cite{Abbott:2016blz,Abbott:2016nmj}
clarified that 
the observed GWs are consistent with the prediction of General Relativity (GR)
for binary coalescence waveforms.
Moreover, the almost simultaneous detection
of GWs from a neutron star (NS) merger \cite{TheLIGOScientific:2017qsa},
and 
the short gamma-ray burst \cite{GBM:2017lvd}
has significantly constrained 
a deviation of propagation speed of GWs over cosmological distance 
from the speed of light down order $10^{-15}$~\cite{Monitor:2017mdv}.  
The future measurements of GWs with unprecedented accuracies
will make it possible to test modified gravity 
from completely different aspects.

Various gravitational theories alternative to GR have been proposed
to explain inflation and/or late-time acceleration of the Universe~\cite{Clifton:2011jh}.
Scalar-tensor theories of gravitation involve
the representative frameworks for modification of GR
such as Horndeski theory~\cite{Horndeski:1974wa} 
(or generalized Galileon~\cite{Nicolis:2008in,Deffayet:2009wt,Deffayet:2009mn,Deffayet:2011gz,Kobayashi:2011nu}), 
and even today sensible construction of scalar-tensor theories have been extensively investigated~\cite{Zumalacarregui:2013pma,Gleyzes:2014dya,Gleyzes:2014qga,Motohashi:2014opa,Langlois:2015cwa,Motohashi:2016ftl,Klein:2016aiq,BenAchour:2016fzp,Motohashi:2017eya,Motohashi:2018pxg}.  
The possible deviations from astrophysical and cosmological predictions in GR 
have been explored as smoking guns of these theories~\cite{Clifton:2011jh,Berti:2015itd,Koyama:2015vza}.

The situation changes abruptly by the recent GW observations.
The constraint on the propagation speed of GWs 
severely restricts theories of modified gravity for the late-time accelerated 
expansion~\cite{Lombriser:2015sxa,Lombriser:2016yzn,Creminelli:2017sry,Sakstein:2017xjx,Ezquiaga:2017ekz,Baker:2017hug} and 
those with the screening mechanism~\cite{Crisostomi:2017lbg,Langlois:2017dyl,Bartolo:2017ibw,Dima:2017pwp}. 
Moreover, the worldwide network of GW interferometer 
will include KAGRA~\cite{Aso:2013eba}, 
and further improve these tests of gravity
both on strong-field regime and cosmological scales.
Within next few years, it is plausible 
that no deviation from predictions in GR would be detected.
If it is the case,
GR or modified gravity theories sharing the same background solutions
and perturbation dynamics with GR 
would be observationally preferred
\footnote{
It should be emphasized that
even if GR and modified gravity theories share the same background solution,
it is not necessarily true that
the perturbation dynamics is also the same in both theories,
as firstly addressed in Ref.~\cite{Barausse:2008xv} for specific theories.
Nevertheless,
our point is that 
if the observational data agree with the predictions of the perturbations in GR,
it would suggest that the background solution
is given by a GR solution.}.

It is then important to note that no detection of deviation from GR predictions 
does not immediately exclude modified gravity theories especially in
strong-field regime, as many theories could share 
the same solutions with GR.
In GR, there is the no-hair theorem which states that the BH spacetime is solely
determined by three conserved quantities or ``hairs'';
mass, angular momentum, and electric charge~\cite{Israel:1967wq,Carter:1971zc,Hawking:1971vc}. 
In general, scalar-tensor theories may possess BH solutions
with nontrivial scalar hair~\cite{Bocharova:1970skc,Bekenstein:1974sf,Kanti:1995vq,Pani:2009wy,Kleihaus:2011tg,Ayzenberg:2014aka,
Sotiriou:2013qea,Anabalon:2013oea,Minamitsuji:2013ura,Erices:2017izj,Babichev:2013cya,Herdeiro:2014goa,Charmousis:2014zaa,Radu:2005bp,Anabalon:2009qt,Kolyvaris:2011fk,Anabalon:2012ta,Babichev:2016fbg,Babichev:2017guv}
which are different from GR BH solutions
with the constant profile of the scalar fields.
Interestingly, however, there exist some class of modified gravity theories
allowing only the BH metric solutions in GR with constant scalar field as the unique solutions~
\cite{Chase,Bekenstein:1972ny,Graham:2014mda,Hawking:1972qk,Bekenstein:1995un,Sotiriou:2011dz,Faraoni:2017ock,Hui:2012qt,Herdeiro:2015waa,Tattersall:2018map}.
This is the extension of no-hair theorems, 
and implies that these classes evade constraints on deviation of BH spacetime from GR one. 
Moreover,
even in a case where GR and non-GR BH solutions exist simultaneously and the
GR BH solution is not the unique solution, if it is the late-time attractor,
the theory dynamically selects the GR BH solution and still evades the constraints.
Therefore, taking into account the rapidly expanding frontier of the modified gravity theories
and the remarkable progress of their constraints from GW observations,
it is important to identify 
which class of the most general scalar-tensor theories could admit GR BH solutions.

In this Letter, we clarify the conditions for the existence of 
GR solutions in a quite general scalar-tensor theory defined by \eqref{action} below,
where by ``GR solution'' we mean a solution 
with a metric satisfying the Einstein equations in GR and a constant profile of the scalar fields.
Our analysis will expand that in Ref.~\cite{Psaltis:2007cw} which showed that 
different gravitational theories share the Kerr solution same as in GR.
Ref.~\cite{Li:2017ncu} constructed the higher-order Ricci polynomial gravity theories
that admit the same vacuum static solutions as GR.
We will cover modified gravity theories 
which can be described by any class of single-/multi-field scalar-tensor theories.
Our analysis solely exploits 
the covariant equations of motion 
without assuming any symmetry and ansatz for the metric and scalar fields, 
and hence any GR solution is within our subject.
Note that ``GR solution''  here
represents not only static or stationary BH solutions such as 
Schwarzschild, Kerr, and Schwarzschild-de Sitter solutions, 
but also any solution in GR 
in astrophysical or cosmological situation
with/without the existence of matter.
Our analysis will also apply higher dimensional spacetime, in which a caveat is that
vacuum GR solutions include not only spherical BHs,
but also black objects with nonspherical horizon topology 
\cite{Emparan:2001wn,Emparan:2008eg}, 
and hence the uniqueness of black objects does not hold.

It should be emphasized that 
our analysis focuses on GR solutions with the constant profile of the scalar fields, 
and there are several theories that do not fit our analysis,
e.g.,
theories with self-gravitating media 
such as Lorentz-violating massive gravity  
\cite{Dubovsky:2004sg,Dubovsky:2005xd,deRham:2010kj,deRham:2011rn,Ballesteros:2016gwc,Celoria:2017bbh},
and 
theories where the small-scale behavior such as breaking of the Vainshtein screening
is sensitive to the asymptotic time-dependence  of the scalar fields
\cite{Kobayashi:2014ida,Koyama:2015oma,Saito:2015fza}.
Correspondingly, 
there are also several examples of 
BH solutions with the metric of GR
in modified gravity theories
that are not captured by the constant scalar field ansatz,
e.g., 
the Schwarzschild-de Sitter BHs  
in the shift-symmetric Horndeski theories \cite{Babichev:2013cya}
and
in the massive gravity theories
\cite{Koyama:2011xz,Koyama:2011yg,Comelli:2011wq,Berezhiani:2011mt,Volkov:2013roa},
and the Kerr solution in the purely quartic Horndeski theory~\cite{Babichev:2017guv}.

\section{The model} \label{sec2}
We consider a wide class of single-/multi-field scalar-tensor theories in $D$-dimensional spacetime described by the action  
\begin{align} 
\label{action} 
S &= \int d^Dx \sqrt{-g}[ G_2(\phi^I,X^{JK}) + G_4 (\phi^I,X^{JK}) R \notag\\
&~~~+ \phi^I_{;\mu_1}C_{1I}^{\mu_1} + \phi^I_{;\mu_1\mu_2}C_{2I}^{\mu_1\mu_2} + \phi^I_{;\mu_1\mu_2\mu_3}C_{3I}^{\mu_1\mu_2\mu_3} + \cdots  \notag\\
&~~~+ L_m(g_{\mu\nu}, \psi) ] , 
\end{align}
where the Greek indices $\mu,\nu,\cdots$ run the $D$-dimensional spacetime,
the capital Latin indices $I,J,\cdots$ label the multiple scalar fields,
and
semicolons denote the covariant derivative with respect to the metric $g_{\mu\nu}$.
In addition to the Ricci curvature $R$ and the matter Lagrangian $L_m(g_{\mu\nu},\psi)$ minimally coupled to gravity, the action involves arbitrary functions: 
$G_2,G_4$ are functions of the multiple scalar fields $\phi^I$ and the kinetic terms $X^{IJ}\equiv -g^{\mu\nu}\phi^I_{;\mu}\phi^J_{;\nu}/2$, 
and 
$C_{1I}^{\mu_1}, C_{2I}^{\mu_1\mu_2}, C_{3I}^{\mu_1\mu_2\mu_3}, \cdots$ are functions of 
$(g_{\alpha\beta},g_{\alpha\beta,\gamma},g_{\alpha\beta,\gamma\delta},\cdots;\phi^I, \phi^I_{;\alpha}, \phi^I_{;\alpha\beta},\cdots ;\epsilon_{\mu\nu\rho\sigma})$
with $\epsilon_{\mu\nu\rho\sigma}$ being the Levi-Civita tensor.
The dots in \eqref{action} contain 
contractions between arbitrary higher-order covariant derivatives of a scalar field 
and its corresponding $C$-function, 
$\phi^I_{;\mu_1\cdots \mu_n}C_{nI}^{\mu_1 \cdots \mu_n}$.
In order for Eq.~\eqref{action} to be covariant with respect to $g_{\mu\nu}$,
the dependence of $C_{1I}^{\mu_1}, C_{2I}^{\mu_1\mu_2}, C_{3I}^{\mu_1\mu_2\mu_3}, \cdots$
on the metric should be through metric itself,  
curvature tensors associated with it, and their covariant derivatives.

This action is very generic and covers a lot of single-/multi-field models of scalar-tensor theories.
Indeed, the term $\phi_{;\mu\nu}C_2^{\mu\nu}$ includes Ostrogradsky ghost-free single-field scalar-tensor theories such as
Horndeski~\cite{Horndeski:1974wa} 
(generalized Galileon~\cite{Nicolis:2008in,Deffayet:2009wt,Deffayet:2009mn,Deffayet:2011gz,Kobayashi:2011nu}), 
Gleyzes-Langlois-Piazza-Vernizzi (GLPV)~\cite{Gleyzes:2014dya,Gleyzes:2014qga}, and 
degenerate higher-order scalar-tensor (DHOST) theories~\cite{Langlois:2015cwa,BenAchour:2016fzp}
as a subclass. 
Specifically, the Horndeski action in the four-dimensional spacetime is described by $C_2^{\mu\nu}=C_{\rm H}^{\mu\nu}$ with
\begin{align} \label{ch}
C_{\rm H}^{\mu\nu} &= G_3g^{\mu\nu} + G_{4X} (g^{\mu\nu}\Box\phi - \phi^{;\mu\nu}) + G_5 G^{\mu\nu} \notag\\
&~~~- \f{1}{6} G_{5X} [g^{\mu\nu}(\Box\phi)^2 - 3\Box\phi \phi^{;\mu\nu} + 2\phi^{;\mu\sigma}{\phi^{;\nu}}_{;\sigma}], 
\end{align}
and GLPV action is given by $C_2^{\mu\nu}=C_{\rm H}^{\mu\nu}+C_{\rm bH}^{\mu\nu}$ with
\begin{align} 
\label{cbh} C_{\rm bH}^{\mu\nu} &= F_4 {\epsilon^{\alpha\beta\mu}}_{\gamma} \epsilon^{\tilde\alpha\tilde\beta\nu\gamma} \phi_{;\alpha}\phi_{;\tilde\alpha} \phi_{;\beta\tilde\beta}  \notag\\
&~~~ + F_5 \epsilon^{\alpha\beta\gamma\mu}\epsilon^{\tilde \alpha\tilde\beta\tilde\gamma\nu} 
\phi_{;\alpha}\phi_{;\tilde \alpha} \phi_{;\beta\tilde\beta} \phi_{;\gamma\tilde\gamma}  ,
\end{align}
where $G_n,F_n$ are functions of $\phi$, 
$X=-g^{\mu\nu}\phi_{;\mu}\phi_{;\nu}/2$, and $G_{nX}\equiv\pa G_n/\pa X$.
Likewise, it is also clear that quadratic- and cubic-order DHOST theories are a subclass and described by the $\phi_{;\mu\nu}C_2^{\mu\nu}$ term.
It also includes parity-violating theories with Chern-Simons term 
or Pontryagin density $\epsilon_{\alpha\beta\gamma\delta}R^{\alpha\beta}{}_{\mu\nu}R^{\gamma\delta\mu\nu}/2$~\cite{Jackiw:2003pm,
Smith:2007jm,
Grumiller:2007rv, 
Yunes:2009hc, 
Konno:2009kg, 
Motohashi:2011pw,Motohashi:2011ds,Ayzenberg:2013wua,Konno:2014qua,Stein:2014xba},
the multi-Galileon theories~\cite{Deffayet:2010zh,Padilla:2010de,Padilla:2010ir,Padilla:2010tj,Trodden:2011xh,Padilla:2012dx,Kobayashi:2013ina,Ohashi:2015fma,Allys:2016hfl},
those with complex scalar fields,
and even 
more general higher-order theories involving derivatives higher than second order, 
which can be free from the Ostrogradsky ghost by imposing a certain set of ghost-free conditions~\cite{Motohashi:2014opa,Motohashi:2017eya,Motohashi:2018pxg}.
Note that in this paper we will focus only on the conditions for obtaining the 
GR solutions 
and actually it does not matter whether the theory \eqref{action} contains the Ostrogradsky ghost or not.
Hence, the following analysis for \eqref{action} to allow GR solutions 
is powerful and exhausts almost all the known scalar-tensor theories of modified gravity.

\section{Conditions for GR solutions}\label{sec3}
We focus on a solution 
in GR with a given value of cosmological constant $\Lambda$ for 
$\Phi^I \equiv (\phi^I,\phi^I_{;\alpha},\phi^I_{;\alpha\beta},\cdots)=\Phi^{I}_0$,
where
$\Phi^I_0\equiv (\phi^I_0,0,0,\cdots)$
and $\phi^I_0$ is constant,
which satisfies the Einstein equation
\begin{align}
\label{gr}
G^{\mu\nu}=8\pi G T^{\mu\nu}-\Lambda g^{\mu\nu},
\end{align}
where 
$T^{\mu\nu} \equiv \f{2}{\sqrt{-g}} \f{\delta (\sqrt{-g} L_m)}{\delta g_{\mu\nu}}$ 
is the stress energy tensor for the matter component,
which is further decomposed
into the classical and constant parts
$T^{\mu\nu}=T_{m}^{\mu\nu} -(8\pi G)^{-1} \Lambda_m g^{\mu\nu}$,
where the latter denotes the contribution of matter vacuum fluctuations.
We elucidate that the theory~\eqref{action} possesses 
GR solutions if the following conditions on the functional forms at 
$\Phi^I= \Phi^I_0$ are satisfied:

\begin{enumerate}

\item \label{con1} 
$G_2, G_4, C_{1I}^{\mu_1}, C_{2I}^{\mu_1\mu_2}, C_{3I}^{\mu_1\mu_2\mu_3}, \cdots$ 
and their derivatives appearing in the Euler-Lagrange equations are regular 
at $\Phi^I= \Phi^I_0$.

\item \label{con2} 
If $T_m^{\mu\nu}=0$,
$8\pi G (G_2 +2\Lambda G_4)=-(16\pi GG_4 -1) \Lambda_m$
and $(D-2)G_{2\phi^I} = -2D (\Lambda+\Lambda_m) G_{4\phi^I}$ 
for $\Phi^I= \Phi^I_0$.
If $T_m^{\mu\nu}\neq 0$, 
$G_4 = (16\pi G)^{-1}$,
$G_2=-\Lambda /(8\pi G)$,
and 
$G_{2\phi^I}=G_{4\phi^I}=0$ 
for $\Phi^I= \Phi^I_0$.

\item \label{con3} 
${C_{1I}^{\mu_1}}_{;\mu_1}={C_{2I}^{\mu_1\mu_2}}_{;\mu_1\mu_2}=\cdots = 0$ for $\Phi^I= \Phi^I_0$.

\end{enumerate}

Since the values of $C_{1I}^{\mu_1}, C_{2I}^{\mu_1\mu_2}, C_{3I}^{\mu_1\mu_2\mu_3} ,\cdots$ 
depend on a choice of the coordinate, in the \hyperref[con1]{condition 1}
we require that as functions of $\Phi^I$ they are regular.
If the \hyperref[con1]{condition 1}  
is not satisfied and some of the functions in the Euler-Lagrange equations diverge at
$\Phi^I= \Phi^I_0$
which occurs e.g.\,, for $G_2\propto \sqrt{X}$ in the cuscuton~\cite{Afshordi:2006ad} 
and more generally $G_n\propto X^{(3-n)/2}$ in the cuscuta-Galileon~\cite{deRham:2016ged}, the divergence of the Euler-Lagrange equations should be avoided either 
by constraining dynamics with nonzero velocity and/or gradient of the scalar field, or 
by cancellation by other divergence with the opposite sign through the entire time evolution. 
An example where the cancellation between singular coupling functions identically holds is  
the Einstein-scalar-Gauss-Bonnet theory defined by 
$L=(\Mpl^2/2)R+X+f(\phi){\cal R}_{\rm GB}$
that is equivalent to the Horndeski theory with
$G_5=-4f'(\phi)\ln X$, 
$G_4=\Mpl^2/2+4f''(\phi) X (2-\ln X)$,
$G_3=-4f^{(3)}(\phi) X (7-3\ln X)$,
and 
$G_2=X+ 8f^{(4)}(\phi) X^2 (3-\ln X)$ \cite{Kobayashi:2011nu}
(with the notation $G_2 \to K$ and $G_3\to -G_3$).
Since $G_{2XX}$, $G_{4X}$, and $C_H^{\mu\nu}$ [See Eq.~\eqref{ch}] and their derivatives
appearing in the Euler-Lagrange equations 
become singular for $X=0$, the \hyperref[con1]{condition 1} is violated.

The \hyperref[con2]{condition 2} depends on the existence of nonzero $T^{\mu\nu}$ and/or $\Lambda$;
for instance, for vacuum solutions in GR with $T_m^{\mu\nu}=\Lambda_m=\Lambda=0$,
the \hyperref[con2]{condition 2} reduces to $G_2=G_{2\phi^I} =0$ 
at $\Phi^I= \Phi^I_0$.

The \hyperref[con3]{condition 3} is satisfied by the Horndeski~\eqref{ch}, GLPV~\eqref{cbh}, 
and DHOST theories so long as their functions $G_3, F_4,\cdots$ and their derivatives are regular 
at $\phi=\phi_0$ and $X=0$.
In contrast, e.g.,\
$C_2^{\mu\nu}\supset R^{\mu\nu}$ violates the \hyperref[con3]{condition 3}
as ${C_2^{\mu\nu}}_{;\mu\nu}\supset g^{\mu\nu} R_{;\mu\nu}/2$ 
from the Bianchi identity, which is nonvanishing in the presence of nonzero matter component. 
Furthermore, 
if $C_2^{\mu\nu}\supset R^{\mu\lambda\rho\sigma}R^{\nu}{}_{\lambda\rho\sigma}$
the \hyperref[con3]{condition 3} is violated even without matter.
The \hyperref[con3]{condition 3} excludes such possibilities.

Since our ``GR solutions'' may be the solutions in GR
with the cosmological constant $\Lambda$ \eqref{gr},
it is manifest that our \hyperref[con1]{conditions 1--3}
are not relevant for the solution of the cosmological constant problem.
If we require it, in addition to the \hyperref[con1]{conditions 1--3},
further fine-tunings for the mass scales would be requested,
which are beyond the scope of our analysis.

Before closing this section, it should be emphasized that 
the \hyperref[con1]{conditions 1--3} are different from those for a no-hair theorem.
In establishing a no-hair theorem (mostly for BH solutions),
the scalar fields $\phi^I$ are assumed to be general functions of the spacetime coordinates
(e.g., functions of the radial coordinate for static and spherically symmetric BH solutions), 
and then the conditions that $\phi^I={\rm const.}$ (and the spacetime metric of a GR solution) is a unique solution are deduced. 
For instance, the no-hair theorem for shift-symmetric Horndeski theory considered in \cite{Hui:2012qt}
adopted a condition for finite Noether current, which amounts to the \hyperref[con1]{condition 1}. 
Under the shift symmetry and the conditions for the Ricci-flat solutions with 
$R^{\mu\nu}=0$,
i.e., $T^{\mu\nu}_{m}=\Lambda_m=\Lambda=0$ from Eq.~\eqref{gr},
our \hyperref[con2]{condition 2} reads $G_2=0$, 
which is consistent with the assumption of the asymptotic flatness
in the no-hair theorem.
The \hyperref[con3]{condition 3} is also identically satisfied.
The uniqueness of GR solution is guaranteed by the additional conditions of 
the staticity, spherical symmetry, and asymptotic flatness of the metric.
In our case
the \hyperref[con1]{conditions 1--3} obtained from the assumptions that
solutions with 
the metric satisfying the Einstein equations in GR
and $\phi^I={\rm const.}$ exist
still allow the non-GR solutions
where the metric is different from any solution in GR
and the scalar fields $\phi^I$ have nontrivial profiles,
and the solutions with the metric of GR but
$\phi^I\neq {\rm const.}$,
as we did not require the uniqueness of the solution. 
Thus, our \hyperref[con1]{conditions 1--3}
should be regarded as the necessary conditions
for establishing a no-hair theorem
when the symmetries of the spacetime 
and the ansatz of the scalar fields are more specified,
e.g., $\phi^I$ are functions of the radial coordinate
for static and spherically symmetric BH spacetimes.

\subsection{Proof}\label{sec3a}
While our statement holds for the wide class of theories~\eqref{action}, as we will see below, the proof is very simple.
We mainly discuss the single-field case of \eqref{action}, as the extension to the multi-field case is straightforward.
We denote the Euler-Lagrange equations for \eqref{action} with respect to the metric and the scalar field as
\begin{align} \label{EOM}
\E^{\mu\nu}&\equiv \f{1}{\sqrt{-g}} \kk{ \f{\pa \L }{\pa g_{\mu\nu}} 
- \pa_\alpha\mk{\f{\pa \L }{\pa g_{\mu\nu,\alpha}}} +\cdots } =0 ,\notag\\
\E_{\phi}&\equiv \f{1}{\sqrt{-g}} \kk{ \f{\pa \L }{\pa \phi} - \nabla_\alpha\mk{\f{\pa \L }{\pa \phi_{;\alpha}} } +\cdots } =0,
\end{align}
which correspond to Einstein equation and Klein-Gordon equation,
and show that for the theories to allow GR solutions, the above 
three conditions should be satisfied.

First, let us focus on the $G_2,G_4$ terms and the matter component $L_m$ in the action~\eqref{action} and set all the $C$-functions zero.  
The Euler-Lagrange equations~\eqref{EOM} are then given by
\begin{align} \label{EOMHorncov}
\E^{\mu\nu}&= \f{1}{2}g^{\mu\nu}G_2 - G^{\mu\nu} G_4 + \f{1}{2}T^{\mu\nu} \notag\\
&~~~- \f{1}{2}(G_{2X}+RG_{4X}) \phi^{;\mu}\phi^{;\nu} + (\nabla^\mu\nabla^\nu-g^{\mu\nu}\Box)G_4 ,\notag\\
\E_\phi&= G_{2\phi} + RG_{4\phi} \notag\\
&~~~+ \f{1}{2}\nabla_\mu (G_{2X}\phi^{;\mu}) + \f{1}{2}R\nabla_\mu(G_{4X}\phi^{;\mu}) .
\end{align}
Substituting $\phi=$~const., 
the second lines of 
$\E^{\mu\nu}$ and $\E_\phi$
vanish so long as we assume the 
\hyperref[con1]{condition 1}, namely, 
$G_2$, $G_4$,
and their derivatives involved in \eqref{EOMHorncov} do not diverge at $\phi=$~const.

Furthermore, substituting \eqref{gr}
and its trace 
$(2-D)R/2=8\pi  G T_m{}^\mu{}_\mu - D (\Lambda+\Lambda_m)$
to \eqref{EOMHorncov}
yields 
\begin{align} \label{EOMHornGRsol}
&g^{\mu\nu}
\left( G_2 +2\Lambda G_4+\frac{16\pi GG_4 -1}{8\pi G}  \Lambda_m \right)
= T_m^{\mu\nu} ( 16\pi G G_4 - 1 ) , \notag
\\
&(D-2)G_{2\phi}
+ 2D( \Lambda+\Lambda_m) G_{4\phi}
= 16 \pi G G_{4\phi} {T_m}{}^\mu{}_{\mu} .
\end{align}
For vacuum solutions, $T_m^{\mu\nu}=0$,
EOMs~\eqref{EOMHornGRsol} match the former case of 
the \hyperref[con2]{condition 2}.
On the other hand, with $T_m^{\mu\nu}\neq 0$, 
each side of \eqref{EOMHornGRsol} has to vanish separately 
as $T_m^{\mu\nu}$ varies with $x^\mu$, leading to the latter case of 
the \hyperref[con2]{condition 2}.
Thus, the action~\eqref{action} without $C$-functions allows  
GR solutions if the 
\hyperref[con1]{condition 1} and the \hyperref[con2]{condition 2}
are satisfied.

Next, we consider the remaining $C$-terms of the action~\eqref{action} and clarify that their contribution to the Euler-Lagrange equations vanish for $\phi=$~const.\
under the \hyperref[con1]{condition 1} and the \hyperref[con3]{condition 3}.
The contribution of $\L_n\equiv \sqrt{-g}\phi_{;\rho_1\cdots\rho_n}C_n^{\rho_1\cdots\rho_n}$ to the Euler-Lagrange equations~\eqref{EOM} is given by
\begin{align} \label{En}
\E^{\mu\nu}&\supset \mk{ \f{1}{2} g^{\mu\nu} C_n^{\rho_1\cdots\rho_n} + \f{\pa C_n^{\rho_1\cdots\rho_n}}{\pa g_{\mu\nu}} } \phi_{;\rho_1\cdots\rho_n} \notag\\
&~~~+ \sum^{\infty}_{k=1} \f{(-1)^k}{\sqrt{-g}} \pa_{\alpha_k} \cdots \pa_{\alpha_1} \mk{  \sqrt{-g} \f{\pa(\phi_{;\rho_1\cdots\rho_n} C_n^{\rho_1\cdots\rho_n}) }{\pa g_{\mu\nu,\alpha_1\cdots \alpha_k}}  } , \notag\\
\E_\phi&\supset \f{\pa C_n^{\rho_1\cdots\rho_n}}{\pa \phi} \phi_{;\rho_1\cdots\rho_n} + (-1)^n {C_n^{\rho_1\cdots\rho_n}}_{;\rho_1\cdots\rho_n} \\
&~~~+ \sum^{\infty}_{k=1} (-1)^k \nabla_{\alpha_k} \cdots \nabla_{\alpha_1} \mk{ \phi_{;\rho_1\cdots\rho_n} \f{\pa C_n^{\rho_1\cdots\rho_n}}{\pa \phi_{;\alpha_1\cdots \alpha_k}} } . \notag
\end{align}
The ${C_n^{\rho_1\cdots\rho_n}}_{;\rho_1\cdots\rho_n}$ term in $\E_\phi$ vanishes from 
the \hyperref[con3]{condition 3}. 
All other terms in the right-hand sides of \eqref{En} are multiplied by derivatives of $\phi$, and thus vanish for $\phi=$~const.\ so long as the $C$-functions and their derivatives are regular.

It should be emphasized that 
our analysis does not include the case that
a GR solution with $\phi=$~const.\ is obtained
via the cancellation of the independent contributions
arising from singular coupling functions
in the Euler-Lagrange equations \eqref{EOM}.
For example, as we mentioned earlier, the Einstein-scalar-Gauss-Bonnet theory 
$L=(\Mpl^2/2)R+X+f(\phi){\cal R}_{\rm GB}$
exhibits such cancellation and 
does not satisfy the \hyperref[con1]{conditions 1--3}.
Nevertheless, it has been shown that the Schwarzschild metric with $\phi=\phi_0={\rm const.}$ 
is a solution if $f'(\phi_0)=0$
\cite{Doneva:2017bvd,Silva:2017uqg,Antoniou:2017acq,Antoniou:2017hxj}.
This is an example of exceptional cases of our proof.

In summary, deriving Euler-Lagrange equations for the full action~\eqref{action} 
and then plugging GR solution $\phi=$~const.\
and \eqref{gr} lead to the equations \eqref{EOMHornGRsol}.
Generalization to the multi-field case is also straightforward. 
We thus conclude that the general theories~\eqref{action} allow 
GR solutions if the 
\hyperref[con1]{conditions 1--3} are satisfied.

\section{Examples}\label{sec4}
Now it is intriguing to consider specific examples.
From the fact that GW observations are consistent with GR, 
we are interested in identifying a class of modified gravity that allows GR solutions
by satisfying the \hyperref[con1]{conditions 1--3}.

Let us consider single-field models in the four-dimensional spacetime.
For the Horndeski theory with \eqref{ch}, the Euler-Lagrange equations
for a static, spherically-symmetric spacetime were derived in \cite{Kobayashi:2012kh}.
In \cite{Kobayashi:2014wsa}, the no-hair vacuum solution with $\Lambda=0$ is considered 
by assuming asymptotic flatness of the spacetime and the scalar field profile $\phi=\phi(r)$, 
and it is clarified that the Euler-Lagrange equations allow the Schwarzschild solution
if \hyperref[con1]{the condition 1} and the \hyperref[con2]{condition 2} are satisfied,
where the \hyperref[con2]{condition 2} reduces to $G_2=G_{2\phi} =0$.
Note that in this case the \hyperref[con3]{condition 3} is automatically satisfied
as long as $G_3$, $G_4$, and $G_5$ are regular for $\phi=$~const.
Note also that the argument in the present paper 
is fully performed in the covariant manner without 
any ansatz for the metric and the scalar field, 
and in general the assumptions such as the asymptotic flatness and $\phi=\phi(r)$
are not necessary.

In the same vein, we can consider GLPV and DHOST theories and 
check that our statement holds.  
We consider a static, spherically-symmetric metric ansatz
\be \label{metric} ds^2 = -A(r)dt^2+\f{dr^2}{B(r)} + r^2(d\theta^2+\sin^2\theta d\varphi^2) , \ee
and plug it to the action~\eqref{action}.
While we fix the gauge as $g_{tr}=0$ and $g_{\theta\theta}=r^2$ at the action level, 
we do not lose any independent EOMs, 
regardless of a choice of a specific theory (see Sec.~V~C of \cite{Motohashi:2016prk}).
EOMs for GLPV theory are given by $\E^{\rm H}_Q+\E^{\rm bH4}_Q+\E^{\rm bH5}_Q=0$ with $Q=A,B,\phi$, 
where the Horndeski terms~$\E^{\rm H}_Q$ are given in \cite{Kobayashi:2012kh,Kobayashi:2014wsa} 
(with the notation $G_2 \to K$ and $G_3\to -G_3$),  
and $\E^{\rm bH4}_Q$ and $\E^{\rm bH5}_Q$ are contributions from the beyond Horndeski terms $F_4$ and $F_5$ in \eqref{cbh}, respectively,
and given by \eqref{Ebh4} and \eqref{Ebh5} in \hyperref[appa]{Appendix A}.
Clearly, assuming that $F_4$, $F_5$, and their derivatives in EOMs are not singular at $\phi=$~const.,
$\E^{\rm bH4}_Q$ and $\E^{\rm bH5}_Q$ vanish for $\phi=$~const. 
Likewise, we explicitly checked that EOMs for quadratic- and cubic-order DHOST theories
vanish for $\phi=$~const.\ so long as we assume the regularity.

In summary, to guarantee the existence of GR solutions, 
the form of $G_2$ and $G_4$ are severely constrained by  
the \hyperref[con2]{condition 2}
and the form of other functions in Horndeski, GLPV, and DHOST theories
are not constrained so long as they are regular at $\phi=$~const.
The condition is different from
the condition $G_4=G_4(\phi)$ and $G_5=$~const.\ obtained in \cite{Creminelli:2017sry,Sakstein:2017xjx,Ezquiaga:2017ekz,Baker:2017hug} 
by imposing the propagation speed of GWs to be the same as GR,
as $G_4$ and $G_5$ are only terms that have 
nonminimal coupling to gravity and affects GW propagation speed. 
Note that our condition and the constraint on $G_4$ and $G_5$ from the propagation speed of GWs 
are independent from several aspects;
the latter is observational, valid only on cosmological scales,
and can be applied to the models where the scalar fields act as the source of dark energy. 
Thus, the models which satisfy the GW constraint on the large scales
would contain non-GR solutions on small scales.

\section{Classification}\label{sec5}
Let us describe further application of the 
\hyperref[con1]{conditions 1--3} 
to general theories of modified gravity on strong-field regime.
The \hyperref[con1]{conditions 1--3} 
classify general theories of modified gravity into three classes.
The classification clarifies the origin of differences between 
many known examples of no-hair theorem and hairy solutions,
and helps us to explore 
GR and non-GR solutions in
theories of modified gravity in various contexts explained below.

First, if a theory satisfies the 
\hyperref[con1]{conditions 1--3}, 
it of course allows GR solutions, 
but may also allow other solutions with a nontrivial scalar field(s).
Therefore, the \hyperref[con1]{conditions 1--3} are necessary conditions for a no-hair theorem, 
and theories satisfying the \hyperref[con1]{conditions 1--3}
serves as a candidate in which a no-hair theorem can be established.
If a GR solution is unique solution, such a theory intrinsically 
passes constraints on deviation from GR.
To guarantee the uniqueness of GR solutions, 
one may need to impose some additional conditions,
for instance, 
symmetries of spacetime,
ansatz for scalar field, 
and/or internal symmetry of the theory.
Indeed, theories satisfying the 
\hyperref[con1]{conditions 1--3} include
Brans-Dicke theory,
the shift-symmetric Horndeski theory, and the shift-symmetric GLPV theory
as a subclass, for which no-hair theorems in the four-dimensional spacetime
have been proven~\cite{Sotiriou:2011dz,Faraoni:2017ock,Hui:2012qt,Lehebel:2017fag}.

On the other hand, if a theory does not satisfy at least 
one of the \hyperref[con1]{conditions 1--3},
it inevitably possesses only non-GR solutions
\footnote{
As we emphasized at the end of 
\hyperref[sec1]{Sec.~I},
throughout the paper we state ``GR solutions'' as solutions satisfying the Einstein equation \eqref{gr} and the constant scalar field profile $\Phi^I=\Phi^I_0$. 
Another type of 
solution satisfying the Einstein equation
for a nonconstant profile of the scalar field,
e.g., the stealth Schwarzschild solution \cite{Babichev:2013cya},
is included in ``non-GR solutions'' here.
}, 
except for the case of
the cancellation discussed in the second paragraph from the last
in \hyperref[sec3]{Sec.~III}.
Therefore, focusing on the violation of the 
\hyperref[con1]{conditions 1--3}, 
one can identify the candidate classes that possess analytic solutions of hairy BH.
An example is
the Einstein-scalar-Gauss-Bonnet theory discussed in~\cite{Sotiriou:2013qea,Kanti:1995vq,Pani:2009wy,Kleihaus:2011tg,Ayzenberg:2014aka} with $G_5\sim \ln X$ in \eqref{ch}, which violates 
the \hyperref[con1]{condition 1}, leading to hairy BH solutions
\footnote{
As we mentioned briefly at the end of 
\hyperref[sec3]{Sec.~III}, 
Refs.~\cite{Doneva:2017bvd,Silva:2017uqg,Antoniou:2017acq,Antoniou:2017hxj}
showed that the Einstein-scalar-Gauss-Bonnet theories admit the Schwarzschild BH solution, 
which would be unstable due to the tachyonic instability triggered by
coupling of the scalar field to the Gauss-Bonnet term.
Since the theory does not satisfy
the \hyperref[con1]{conditions 1--3},
the realization of the Schwarzschild solution 
exploits the cancellation of the terms absent in GR,
which falls in the exception of our analysis as mentioned above.
}.
Other examples of hairy BH solutions 
obtained by the violation of the \hyperref[con1]{condition 1} 
were explicitly constructed in~\cite{Babichev:2017guv}
for the Horndeski and beyond-Horndeski theories
in which $G_i$ and $F_i$ in \eqref{ch} and \eqref{cbh} 
and their derivatives are not analytic at $X=0$.

The last possibility is inbetween the above two:
a class that possesses GR solutions by satisfying the 
\hyperref[con1]{conditions 1--3} 
and allows other hairy solutions at the same time, 
one of which may be attractor. 
If a GR solution is the attractor, 
it may be said that a no-hair theorem holds in a dynamical way, 
and it passes observational constraints on deviation from GR spacetime.
In contrast, the opposite case 
that a non-GR BH solution is dynamically selected
rather than a GR one could also happen, such as
the spontaneous scalarization~\cite{Damour:1993hw,Damour:1996ke}.
The key for the coexistence of GR and non-GR solutions is that 
the theory satisfies the \hyperref[con1]{conditions 1--3} to allow GR solutions, 
and also has some internal symmetry to allow hairy BH solutions
without spoiling spacetime symmetry.
Specific examples of this class include
a hairy solution in a subclass of the shift-symmetric Horndeski theories with 
a nontrivial linear time dependence of the scalar field $\phi=qt+\psi(r)$~\cite{Babichev:2013cya}
(see also \cite{Kobayashi:2014eva}),
the Kerr-like hairy solution in Einstein-complex scalar theory with $U(1)$ symmetry 
with the complex scalar field profile $\Phi=\phi(r,\theta)e^{i(m\varphi-\omega t)}$~\cite{Herdeiro:2014goa},
the Bocharova-Bronnikov-Melnikov-Bekenstein (BBMB) solution \cite{Bocharova:1970skc,Bekenstein:1974sf} in a conformally coupled scalar field,
and a similar solution in a two-field extension of the Horndeski theory~\cite{Charmousis:2014zaa}. 
Checking whether a GR or non-GR solution is the attractor requires further studies on a case-by-case basis.

\section{Conclusion} \label{sec6}
The recent and future GW observations allow us to place a stringent constraint on deviation from GR, 
and hence it is important to identify theories of modified gravity 
that can intrinsically share the same solutions with GR.
We have investigated a quite general class~\eqref{action} of single-/multi-field scalar-tensor theories 
with arbitrary higher-order derivatives in arbitrary spacetime dimension.
We confirmed that GR solutions
are allowed if the \hyperref[con1]{conditions 1--3} are satisfied.
This approach yields an independent result from
the one requiring 
the propagation speed of gravitational wave to be the speed of light
as in GR \cite{GBM:2017lvd}.
Our analysis was fully covariant, and hence can be applied to any astrophysical or cosmological situation.
The \hyperref[con1]{conditions 1--3} classify general theories of modified gravity into three classes, 
each of which possesses 
i) only GR solutions (i.e., no-hair cases), 
ii) only hairy solutions (except the cases that GR solutions are realized by cancellation between singular coupling functions in the Euler-Lagrange equations), 
and 
iii) both GR and hairy solutions, 
for the last of which one of the two solutions may be selected dynamically.

There will be several extensions of our analysis.
One of them is the possibility of GR solutions with $X={\rm const}$.
The simplest example is the Schwarzschild solution
with a nontrivial scalar field \cite{Babichev:2013cya},
and it would be important to see whether theories \eqref{action}
also admit generic GR solutions with $X={\rm const.}$ and clarify characteristics in their dynamics.

Furthermore, while we have focused on the scalar-tensor theories,
it would be definitively important to extend our analysis
to the theories which include vector fields, fermions, and other tensor fields as in bigravity theories. 
In such higher-spin theories
the conditions that they admit GR solutions would be able to be obtained 
by requesting that all components of vector, fermion and tensor fields vanish
in the entire spacetime. 
More rigorous derivation of the conditions for each theory 
is out of scope of our paper and will be left for the future work.

\acknowledgments{
We thank V.\ Cardoso and S.\ Mukohyama for useful comments on the manuscript.
H.M.\ was supported in part by JSPS KEKENHI Grant No.\ JP17H06359.
M.M.\ was supported by FCT-Portugal through Grant No.\ SFRH/BPD/88299/2012. 
H.M.\ acknowledges CENTRA for hospitality, where part of this work was completed.
}

\appendix

\section{EOMs for beyond Horndeski Lagrangians} \label{appa}

EOMs for the beyond Horndeski terms $F_4$ and $F_5$ in \eqref{cbh} 
for the static, spherically-symmetric metric ansatz~\eqref{metric} are respectively given by
\begin{align} \label{Ebh4}
&\E^{\rm bH4}_A = -\f{2B^3\phi'^3}{r^2} \left[ \mk{ \phi' + 5\f{B'}{B} r \phi'+ 8 r \phi''} F_4 \right.\notag\\
& \left. + 2r\phi'^2F_{4\phi} - r\phi' (B\phi'^2)'F_{4X} \right]  ,\notag\\ 
&\E^{\rm bH4}_B = \f{2AB^2\phi'^4}{r^2} \f{(rA)'}{A} (5F_4-B\phi'^2F_{4X}) , \\ 
&\E^{\rm bH4}_\phi = \f{2 A B^3\phi'^2 }{r^2} \left[ 4 \left\{  \mk{ \alpha - 5 \f{A'}{A} \f{B'}{B} } r \phi' \right. \right. \notag\\
&\left. - \mk{ 3\f{A'}{A} + 5 \f{B'}{B} } \phi' - 6 \f{(rA)'}{A} \phi''  \right\}  F_4  
+ \left\{ 18 \f{(rA)'}{A} \phi'' \right.\notag\\
& \left. + \mk{ - \alpha + 11\f{A'}{A} \f{B'}{B} } r \phi' + \mk{ 3 \f{A'}{A} + 11 \f{B'}{B}} \phi' \right\} B \phi'^2 F_{4X}  \notag\\
& \left. - \f{(rA)'}{A} \phi'^2 \ck{ 6 F_{4\phi} - 2 B \phi'^2 F_{4\phi X} + (B\phi'^2)' B \phi' F_{4XX} } \right] , \notag
\end{align}
and
\begin{align} \label{Ebh5} 
&\E^{\rm bH5}_A = -\f{3B^4\phi'^4}{r^2} \left[ \mk{ \f{7B'}{B}\phi' + 10\phi'' } F_5 \right.\notag\\
& \left. + 2\phi'^2 F_{5\phi} - \phi' (B\phi'^2)' F_{5X} \right] ,\notag\\ 
&\E^{\rm bH5}_B = \f{3A'B^3\phi'^5}{r^2} (7F_5-B\phi'^2F_{5X}) ,\\ 
&\E^{\rm bH5}_\phi = \f{3A B^4 \phi'^3}{r^2} \left[ 5 \left\{ \mk{ \alpha -7\f{A'}{A} \f{B'}{B}} \phi'  
- 8 \f{A'}{A} \phi'' \right\} F_5 \right. \notag\\
& + \ck{ \mk{ - \alpha + 14\f{A'}{A}\f{B'}{B}} \phi' + 22 \f{A'}{A} \phi''} B \phi'^2 F_{5X} \notag\\
&  \left. - \f{A'}{A} \phi'^2 \ck{ 8  F_{5\phi} - 2 B \phi'^2 F_{5\phi X} + (B\phi'^2)'  B \phi' F_{5XX} } \right], \notag
\end{align}
where $\alpha\equiv \f{A'^2}{A^2} - 2 \f{A''}{A}$.

\bibliography{ref-nohair}

\begin{thebibliography}{118}%
\makeatletter
\providecommand \@ifxundefined [1]{%
 \@ifx{#1\undefined}
}%
\providecommand \@ifnum [1]{%
 \ifnum #1\expandafter \@firstoftwo
 \else \expandafter \@secondoftwo
 \fi
}%
\providecommand \@ifx [1]{%
 \ifx #1\expandafter \@firstoftwo
 \else \expandafter \@secondoftwo
 \fi
}%
\providecommand \natexlab [1]{#1}%
\providecommand \enquote  [1]{``#1''}%
\providecommand \bibnamefont  [1]{#1}%
\providecommand \bibfnamefont [1]{#1}%
\providecommand \citenamefont [1]{#1}%
\providecommand \href@noop [0]{\@secondoftwo}%
\providecommand \href [0]{\begingroup \@sanitize@url \@href}%
\providecommand \@href[1]{\@@startlink{#1}\@@href}%
\providecommand \@@href[1]{\endgroup#1\@@endlink}%
\providecommand \@sanitize@url [0]{\catcode `\\12\catcode `\$12\catcode
  `\&12\catcode `\#12\catcode `\^12\catcode `\_12\catcode `\%12\relax}%
\providecommand \@@startlink[1]{}%
\providecommand \@@endlink[0]{}%
\providecommand \url  [0]{\begingroup\@sanitize@url \@url }%
\providecommand \@url [1]{\endgroup\@href {#1}{\urlprefix }}%
\providecommand \urlprefix  [0]{URL }%
\providecommand \Eprint [0]{\href }%
\providecommand \doibase [0]{http://dx.doi.org/}%
\providecommand \selectlanguage [0]{\@gobble}%
\providecommand \bibinfo  [0]{\@secondoftwo}%
\providecommand \bibfield  [0]{\@secondoftwo}%
\providecommand \translation [1]{[#1]}%
\providecommand \BibitemOpen [0]{}%
\providecommand \bibitemStop [0]{}%
\providecommand \bibitemNoStop [0]{.\EOS\space}%
\providecommand \EOS [0]{\spacefactor3000\relax}%
\providecommand \BibitemShut  [1]{\csname bibitem#1\endcsname}%
\let\auto@bib@innerbib\@empty
\bibitem [{\citenamefont {Abbott}\ \emph
  {et~al.}(2016{\natexlab{a}})\citenamefont {Abbott} \emph
  {et~al.}}]{Abbott:2016blz}%
  \BibitemOpen
  \bibfield  {author} {\bibinfo {author} {\bibfnamefont {B.~P.}\ \bibnamefont
  {Abbott}} \emph {et~al.} (\bibinfo {collaboration} {Virgo, LIGO
  Scientific}),\ }\href {\doibase 10.1103/PhysRevLett.116.061102} {\bibfield
  {journal} {\bibinfo  {journal} {Phys. Rev. Lett.}\ }\textbf {\bibinfo
  {volume} {116}},\ \bibinfo {pages} {061102} (\bibinfo {year}
  {2016}{\natexlab{a}})},\ \Eprint {http://arxiv.org/abs/1602.03837}
  {arXiv:1602.03837 [gr-qc]} \BibitemShut {NoStop}%
\bibitem [{\citenamefont {Abbott}\ \emph
  {et~al.}(2016{\natexlab{b}})\citenamefont {Abbott} \emph
  {et~al.}}]{Abbott:2016nmj}%
  \BibitemOpen
  \bibfield  {author} {\bibinfo {author} {\bibfnamefont {B.~P.}\ \bibnamefont
  {Abbott}} \emph {et~al.} (\bibinfo {collaboration} {Virgo, LIGO
  Scientific}),\ }\href {\doibase 10.1103/PhysRevLett.116.241103} {\bibfield
  {journal} {\bibinfo  {journal} {Phys. Rev. Lett.}\ }\textbf {\bibinfo
  {volume} {116}},\ \bibinfo {pages} {241103} (\bibinfo {year}
  {2016}{\natexlab{b}})},\ \Eprint {http://arxiv.org/abs/1606.04855}
  {arXiv:1606.04855 [gr-qc]} \BibitemShut {NoStop}%
\bibitem [{\citenamefont {Abbott}\ \emph
  {et~al.}(2017{\natexlab{a}})\citenamefont {Abbott} \emph
  {et~al.}}]{TheLIGOScientific:2017qsa}%
  \BibitemOpen
  \bibfield  {author} {\bibinfo {author} {\bibfnamefont {B.}~\bibnamefont
  {Abbott}} \emph {et~al.} (\bibinfo {collaboration} {Virgo, LIGO
  Scientific}),\ }\href {\doibase 10.1103/PhysRevLett.119.161101} {\bibfield
  {journal} {\bibinfo  {journal} {Phys. Rev. Lett.}\ }\textbf {\bibinfo
  {volume} {119}},\ \bibinfo {pages} {161101} (\bibinfo {year}
  {2017}{\natexlab{a}})},\ \Eprint {http://arxiv.org/abs/1710.05832}
  {arXiv:1710.05832 [gr-qc]} \BibitemShut {NoStop}%
\bibitem [{\citenamefont {Abbott}\ \emph
  {et~al.}(2017{\natexlab{b}})\citenamefont {Abbott} \emph
  {et~al.}}]{GBM:2017lvd}%
  \BibitemOpen
  \bibfield  {author} {\bibinfo {author} {\bibfnamefont {B.~P.}\ \bibnamefont
  {Abbott}} \emph {et~al.} (\bibinfo {collaboration} {GROND, SALT Group,
  OzGrav, DFN, INTEGRAL, Virgo, Insight-Hxmt, MAXI Team, Fermi-LAT, J-GEM,
  RATIR, IceCube, CAASTRO, LWA, ePESSTO, GRAWITA, RIMAS, SKA South
  Africa/MeerKAT, H.E.S.S., 1M2H Team, IKI-GW Follow-up, Fermi GBM, Pi of Sky,
  DWF (Deeper Wider Faster Program), Dark Energy Survey, MASTER, AstroSat
  Cadmium Zinc Telluride Imager Team, Swift, Pierre Auger, ASKAP, VINROUGE,
  JAGWAR, Chandra Team at McGill University, TTU-NRAO, GROWTH, AGILE Team, MWA,
  ATCA, AST3, TOROS, Pan-STARRS, NuSTAR, ATLAS Telescopes, BOOTES, CaltechNRAO,
  LIGO Scientific, High Time Resolution Universe Survey, Nordic Optical
  Telescope, Las Cumbres Observatory Group, TZAC Consortium, LOFAR, IPN, DLT40,
  Texas Tech University, HAWC, ANTARES, KU, Dark Energy Camera GW-EM, CALET,
  Euro VLBI Team, ALMA}),\ }\href {\doibase 10.3847/2041-8213/aa91c9}
  {\bibfield  {journal} {\bibinfo  {journal} {Astrophys. J.}\ }\textbf
  {\bibinfo {volume} {848}},\ \bibinfo {pages} {L12} (\bibinfo {year}
  {2017}{\natexlab{b}})},\ \Eprint {http://arxiv.org/abs/1710.05833}
  {arXiv:1710.05833 [astro-ph.HE]} \BibitemShut {NoStop}%
\bibitem [{\citenamefont {Abbott}\ \emph
  {et~al.}(2017{\natexlab{c}})\citenamefont {Abbott} \emph
  {et~al.}}]{Monitor:2017mdv}%
  \BibitemOpen
  \bibfield  {author} {\bibinfo {author} {\bibfnamefont {B.~P.}\ \bibnamefont
  {Abbott}} \emph {et~al.} (\bibinfo {collaboration} {Virgo, Fermi-GBM,
  INTEGRAL, LIGO Scientific}),\ }\href {\doibase 10.3847/2041-8213/aa920c}
  {\bibfield  {journal} {\bibinfo  {journal} {Astrophys. J.}\ }\textbf
  {\bibinfo {volume} {848}},\ \bibinfo {pages} {L13} (\bibinfo {year}
  {2017}{\natexlab{c}})},\ \Eprint {http://arxiv.org/abs/1710.05834}
  {arXiv:1710.05834 [astro-ph.HE]} \BibitemShut {NoStop}%
\bibitem [{\citenamefont {Clifton}\ \emph {et~al.}(2012)\citenamefont
  {Clifton}, \citenamefont {Ferreira}, \citenamefont {Padilla},\ and\
  \citenamefont {Skordis}}]{Clifton:2011jh}%
  \BibitemOpen
  \bibfield  {author} {\bibinfo {author} {\bibfnamefont {T.}~\bibnamefont
  {Clifton}}, \bibinfo {author} {\bibfnamefont {P.~G.}\ \bibnamefont
  {Ferreira}}, \bibinfo {author} {\bibfnamefont {A.}~\bibnamefont {Padilla}}, \
  and\ \bibinfo {author} {\bibfnamefont {C.}~\bibnamefont {Skordis}},\ }\href
  {\doibase 10.1016/j.physrep.2012.01.001} {\bibfield  {journal} {\bibinfo
  {journal} {Phys. Rept.}\ }\textbf {\bibinfo {volume} {513}},\ \bibinfo
  {pages} {1} (\bibinfo {year} {2012})},\ \Eprint
  {http://arxiv.org/abs/1106.2476} {arXiv:1106.2476 [astro-ph.CO]} \BibitemShut
  {NoStop}%
\bibitem [{\citenamefont {Horndeski}(1974)}]{Horndeski:1974wa}%
  \BibitemOpen
  \bibfield  {author} {\bibinfo {author} {\bibfnamefont {G.~W.}\ \bibnamefont
  {Horndeski}},\ }\href {\doibase 10.1007/BF01807638} {\bibfield  {journal}
  {\bibinfo  {journal} {Int. J. Theor. Phys.}\ }\textbf {\bibinfo {volume}
  {10}},\ \bibinfo {pages} {363} (\bibinfo {year} {1974})}\BibitemShut
  {NoStop}%
\bibitem [{\citenamefont {Nicolis}\ \emph {et~al.}(2009)\citenamefont
  {Nicolis}, \citenamefont {Rattazzi},\ and\ \citenamefont
  {Trincherini}}]{Nicolis:2008in}%
  \BibitemOpen
  \bibfield  {author} {\bibinfo {author} {\bibfnamefont {A.}~\bibnamefont
  {Nicolis}}, \bibinfo {author} {\bibfnamefont {R.}~\bibnamefont {Rattazzi}}, \
  and\ \bibinfo {author} {\bibfnamefont {E.}~\bibnamefont {Trincherini}},\
  }\href {\doibase 10.1103/PhysRevD.79.064036} {\bibfield  {journal} {\bibinfo
  {journal} {Phys. Rev.}\ }\textbf {\bibinfo {volume} {D79}},\ \bibinfo {pages}
  {064036} (\bibinfo {year} {2009})},\ \Eprint {http://arxiv.org/abs/0811.2197}
  {arXiv:0811.2197 [hep-th]} \BibitemShut {NoStop}%
\bibitem [{\citenamefont {Deffayet}\ \emph
  {et~al.}(2009{\natexlab{a}})\citenamefont {Deffayet}, \citenamefont
  {Esposito-Farese},\ and\ \citenamefont {Vikman}}]{Deffayet:2009wt}%
  \BibitemOpen
  \bibfield  {author} {\bibinfo {author} {\bibfnamefont {C.}~\bibnamefont
  {Deffayet}}, \bibinfo {author} {\bibfnamefont {G.}~\bibnamefont
  {Esposito-Farese}}, \ and\ \bibinfo {author} {\bibfnamefont {A.}~\bibnamefont
  {Vikman}},\ }\href {\doibase 10.1103/PhysRevD.79.084003} {\bibfield
  {journal} {\bibinfo  {journal} {Phys. Rev.}\ }\textbf {\bibinfo {volume}
  {D79}},\ \bibinfo {pages} {084003} (\bibinfo {year} {2009}{\natexlab{a}})},\
  \Eprint {http://arxiv.org/abs/0901.1314} {arXiv:0901.1314 [hep-th]}
  \BibitemShut {NoStop}%
\bibitem [{\citenamefont {Deffayet}\ \emph
  {et~al.}(2009{\natexlab{b}})\citenamefont {Deffayet}, \citenamefont {Deser},\
  and\ \citenamefont {Esposito-Farese}}]{Deffayet:2009mn}%
  \BibitemOpen
  \bibfield  {author} {\bibinfo {author} {\bibfnamefont {C.}~\bibnamefont
  {Deffayet}}, \bibinfo {author} {\bibfnamefont {S.}~\bibnamefont {Deser}}, \
  and\ \bibinfo {author} {\bibfnamefont {G.}~\bibnamefont {Esposito-Farese}},\
  }\href {\doibase 10.1103/PhysRevD.80.064015} {\bibfield  {journal} {\bibinfo
  {journal} {Phys. Rev.}\ }\textbf {\bibinfo {volume} {D80}},\ \bibinfo {pages}
  {064015} (\bibinfo {year} {2009}{\natexlab{b}})},\ \Eprint
  {http://arxiv.org/abs/0906.1967} {arXiv:0906.1967 [gr-qc]} \BibitemShut
  {NoStop}%
\bibitem [{\citenamefont {Deffayet}\ \emph {et~al.}(2011)\citenamefont
  {Deffayet}, \citenamefont {Gao}, \citenamefont {Steer},\ and\ \citenamefont
  {Zahariade}}]{Deffayet:2011gz}%
  \BibitemOpen
  \bibfield  {author} {\bibinfo {author} {\bibfnamefont {C.}~\bibnamefont
  {Deffayet}}, \bibinfo {author} {\bibfnamefont {X.}~\bibnamefont {Gao}},
  \bibinfo {author} {\bibfnamefont {D.~A.}\ \bibnamefont {Steer}}, \ and\
  \bibinfo {author} {\bibfnamefont {G.}~\bibnamefont {Zahariade}},\ }\href
  {\doibase 10.1103/PhysRevD.84.064039} {\bibfield  {journal} {\bibinfo
  {journal} {Phys. Rev.}\ }\textbf {\bibinfo {volume} {D84}},\ \bibinfo {pages}
  {064039} (\bibinfo {year} {2011})},\ \Eprint {http://arxiv.org/abs/1103.3260}
  {arXiv:1103.3260 [hep-th]} \BibitemShut {NoStop}%
\bibitem [{\citenamefont {Kobayashi}\ \emph {et~al.}(2011)\citenamefont
  {Kobayashi}, \citenamefont {Yamaguchi},\ and\ \citenamefont
  {Yokoyama}}]{Kobayashi:2011nu}%
  \BibitemOpen
  \bibfield  {author} {\bibinfo {author} {\bibfnamefont {T.}~\bibnamefont
  {Kobayashi}}, \bibinfo {author} {\bibfnamefont {M.}~\bibnamefont
  {Yamaguchi}}, \ and\ \bibinfo {author} {\bibfnamefont {J.}~\bibnamefont
  {Yokoyama}},\ }\href {\doibase 10.1143/PTP.126.511} {\bibfield  {journal}
  {\bibinfo  {journal} {Prog. Theor. Phys.}\ }\textbf {\bibinfo {volume}
  {126}},\ \bibinfo {pages} {511} (\bibinfo {year} {2011})},\ \Eprint
  {http://arxiv.org/abs/1105.5723} {arXiv:1105.5723 [hep-th]} \BibitemShut
  {NoStop}%
\bibitem [{\citenamefont {Zumalacárregui}\ and\ \citenamefont
  {García-Bellido}(2014)}]{Zumalacarregui:2013pma}%
  \BibitemOpen
  \bibfield  {author} {\bibinfo {author} {\bibfnamefont {M.}~\bibnamefont
  {Zumalacárregui}}\ and\ \bibinfo {author} {\bibfnamefont {J.}~\bibnamefont
  {García-Bellido}},\ }\href {\doibase 10.1103/PhysRevD.89.064046} {\bibfield
  {journal} {\bibinfo  {journal} {Phys. Rev.}\ }\textbf {\bibinfo {volume}
  {D89}},\ \bibinfo {pages} {064046} (\bibinfo {year} {2014})},\ \Eprint
  {http://arxiv.org/abs/1308.4685} {arXiv:1308.4685 [gr-qc]} \BibitemShut
  {NoStop}%
\bibitem [{\citenamefont {Gleyzes}\ \emph
  {et~al.}(2015{\natexlab{a}})\citenamefont {Gleyzes}, \citenamefont
  {Langlois}, \citenamefont {Piazza},\ and\ \citenamefont
  {Vernizzi}}]{Gleyzes:2014dya}%
  \BibitemOpen
  \bibfield  {author} {\bibinfo {author} {\bibfnamefont {J.}~\bibnamefont
  {Gleyzes}}, \bibinfo {author} {\bibfnamefont {D.}~\bibnamefont {Langlois}},
  \bibinfo {author} {\bibfnamefont {F.}~\bibnamefont {Piazza}}, \ and\ \bibinfo
  {author} {\bibfnamefont {F.}~\bibnamefont {Vernizzi}},\ }\href {\doibase
  10.1103/PhysRevLett.114.211101} {\bibfield  {journal} {\bibinfo  {journal}
  {Phys. Rev. Lett.}\ }\textbf {\bibinfo {volume} {114}},\ \bibinfo {pages}
  {211101} (\bibinfo {year} {2015}{\natexlab{a}})},\ \Eprint
  {http://arxiv.org/abs/1404.6495} {arXiv:1404.6495 [hep-th]} \BibitemShut
  {NoStop}%
\bibitem [{\citenamefont {Gleyzes}\ \emph
  {et~al.}(2015{\natexlab{b}})\citenamefont {Gleyzes}, \citenamefont
  {Langlois}, \citenamefont {Piazza},\ and\ \citenamefont
  {Vernizzi}}]{Gleyzes:2014qga}%
  \BibitemOpen
  \bibfield  {author} {\bibinfo {author} {\bibfnamefont {J.}~\bibnamefont
  {Gleyzes}}, \bibinfo {author} {\bibfnamefont {D.}~\bibnamefont {Langlois}},
  \bibinfo {author} {\bibfnamefont {F.}~\bibnamefont {Piazza}}, \ and\ \bibinfo
  {author} {\bibfnamefont {F.}~\bibnamefont {Vernizzi}},\ }\href {\doibase
  10.1088/1475-7516/2015/02/018} {\bibfield  {journal} {\bibinfo  {journal}
  {JCAP}\ }\textbf {\bibinfo {volume} {1502}},\ \bibinfo {pages} {018}
  (\bibinfo {year} {2015}{\natexlab{b}})},\ \Eprint
  {http://arxiv.org/abs/1408.1952} {arXiv:1408.1952 [astro-ph.CO]} \BibitemShut
  {NoStop}%
\bibitem [{\citenamefont {Motohashi}\ and\ \citenamefont
  {Suyama}(2015)}]{Motohashi:2014opa}%
  \BibitemOpen
  \bibfield  {author} {\bibinfo {author} {\bibfnamefont {H.}~\bibnamefont
  {Motohashi}}\ and\ \bibinfo {author} {\bibfnamefont {T.}~\bibnamefont
  {Suyama}},\ }\href {\doibase 10.1103/PhysRevD.91.085009} {\bibfield
  {journal} {\bibinfo  {journal} {Phys. Rev.}\ }\textbf {\bibinfo {volume}
  {D91}},\ \bibinfo {pages} {085009} (\bibinfo {year} {2015})},\ \Eprint
  {http://arxiv.org/abs/1411.3721} {arXiv:1411.3721 [physics.class-ph]}
  \BibitemShut {NoStop}%
\bibitem [{\citenamefont {Langlois}\ and\ \citenamefont
  {Noui}(2016)}]{Langlois:2015cwa}%
  \BibitemOpen
  \bibfield  {author} {\bibinfo {author} {\bibfnamefont {D.}~\bibnamefont
  {Langlois}}\ and\ \bibinfo {author} {\bibfnamefont {K.}~\bibnamefont
  {Noui}},\ }\href {\doibase 10.1088/1475-7516/2016/02/034} {\bibfield
  {journal} {\bibinfo  {journal} {JCAP}\ }\textbf {\bibinfo {volume} {1602}},\
  \bibinfo {pages} {034} (\bibinfo {year} {2016})},\ \Eprint
  {http://arxiv.org/abs/1510.06930} {arXiv:1510.06930 [gr-qc]} \BibitemShut
  {NoStop}%
\bibitem [{\citenamefont {Motohashi}\ \emph
  {et~al.}(2016{\natexlab{a}})\citenamefont {Motohashi}, \citenamefont {Noui},
  \citenamefont {Suyama}, \citenamefont {Yamaguchi},\ and\ \citenamefont
  {Langlois}}]{Motohashi:2016ftl}%
  \BibitemOpen
  \bibfield  {author} {\bibinfo {author} {\bibfnamefont {H.}~\bibnamefont
  {Motohashi}}, \bibinfo {author} {\bibfnamefont {K.}~\bibnamefont {Noui}},
  \bibinfo {author} {\bibfnamefont {T.}~\bibnamefont {Suyama}}, \bibinfo
  {author} {\bibfnamefont {M.}~\bibnamefont {Yamaguchi}}, \ and\ \bibinfo
  {author} {\bibfnamefont {D.}~\bibnamefont {Langlois}},\ }\href {\doibase
  10.1088/1475-7516/2016/07/033} {\bibfield  {journal} {\bibinfo  {journal}
  {JCAP}\ }\textbf {\bibinfo {volume} {1607}},\ \bibinfo {pages} {033}
  (\bibinfo {year} {2016}{\natexlab{a}})},\ \Eprint
  {http://arxiv.org/abs/1603.09355} {arXiv:1603.09355 [hep-th]} \BibitemShut
  {NoStop}%
\bibitem [{\citenamefont {Klein}\ and\ \citenamefont
  {Roest}(2016)}]{Klein:2016aiq}%
  \BibitemOpen
  \bibfield  {author} {\bibinfo {author} {\bibfnamefont {R.}~\bibnamefont
  {Klein}}\ and\ \bibinfo {author} {\bibfnamefont {D.}~\bibnamefont {Roest}},\
  }\href {\doibase 10.1007/JHEP07(2016)130} {\bibfield  {journal} {\bibinfo
  {journal} {JHEP}\ }\textbf {\bibinfo {volume} {07}},\ \bibinfo {pages} {130}
  (\bibinfo {year} {2016})},\ \Eprint {http://arxiv.org/abs/1604.01719}
  {arXiv:1604.01719 [hep-th]} \BibitemShut {NoStop}%
\bibitem [{\citenamefont {Ben~Achour}\ \emph {et~al.}(2016)\citenamefont
  {Ben~Achour}, \citenamefont {Crisostomi}, \citenamefont {Koyama},
  \citenamefont {Langlois}, \citenamefont {Noui},\ and\ \citenamefont
  {Tasinato}}]{BenAchour:2016fzp}%
  \BibitemOpen
  \bibfield  {author} {\bibinfo {author} {\bibfnamefont {J.}~\bibnamefont
  {Ben~Achour}}, \bibinfo {author} {\bibfnamefont {M.}~\bibnamefont
  {Crisostomi}}, \bibinfo {author} {\bibfnamefont {K.}~\bibnamefont {Koyama}},
  \bibinfo {author} {\bibfnamefont {D.}~\bibnamefont {Langlois}}, \bibinfo
  {author} {\bibfnamefont {K.}~\bibnamefont {Noui}}, \ and\ \bibinfo {author}
  {\bibfnamefont {G.}~\bibnamefont {Tasinato}},\ }\href {\doibase
  10.1007/JHEP12(2016)100} {\bibfield  {journal} {\bibinfo  {journal} {JHEP}\
  }\textbf {\bibinfo {volume} {12}},\ \bibinfo {pages} {100} (\bibinfo {year}
  {2016})},\ \Eprint {http://arxiv.org/abs/1608.08135} {arXiv:1608.08135
  [hep-th]} \BibitemShut {NoStop}%
\bibitem [{\citenamefont {Motohashi}\ \emph
  {et~al.}(2018{\natexlab{a}})\citenamefont {Motohashi}, \citenamefont
  {Suyama},\ and\ \citenamefont {Yamaguchi}}]{Motohashi:2017eya}%
  \BibitemOpen
  \bibfield  {author} {\bibinfo {author} {\bibfnamefont {H.}~\bibnamefont
  {Motohashi}}, \bibinfo {author} {\bibfnamefont {T.}~\bibnamefont {Suyama}}, \
  and\ \bibinfo {author} {\bibfnamefont {M.}~\bibnamefont {Yamaguchi}},\ }\href
  {\doibase 10.7566/JPSJ.87.063401} {\bibfield  {journal} {\bibinfo  {journal}
  {J. Phys. Soc. Jap.}\ }\textbf {\bibinfo {volume} {87}},\ \bibinfo {pages}
  {063401} (\bibinfo {year} {2018}{\natexlab{a}})},\ \Eprint
  {http://arxiv.org/abs/1711.08125} {arXiv:1711.08125 [hep-th]} \BibitemShut
  {NoStop}%
\bibitem [{\citenamefont {Motohashi}\ \emph
  {et~al.}(2018{\natexlab{b}})\citenamefont {Motohashi}, \citenamefont
  {Suyama},\ and\ \citenamefont {Yamaguchi}}]{Motohashi:2018pxg}%
  \BibitemOpen
  \bibfield  {author} {\bibinfo {author} {\bibfnamefont {H.}~\bibnamefont
  {Motohashi}}, \bibinfo {author} {\bibfnamefont {T.}~\bibnamefont {Suyama}}, \
  and\ \bibinfo {author} {\bibfnamefont {M.}~\bibnamefont {Yamaguchi}},\
  }\href@noop {} {\  (\bibinfo {year} {2018}{\natexlab{b}})},\ \Eprint
  {http://arxiv.org/abs/1804.07990} {arXiv:1804.07990 [hep-th]} \BibitemShut
  {NoStop}%
\bibitem [{\citenamefont {Berti}\ \emph {et~al.}(2015)\citenamefont {Berti}
  \emph {et~al.}}]{Berti:2015itd}%
  \BibitemOpen
  \bibfield  {author} {\bibinfo {author} {\bibfnamefont {E.}~\bibnamefont
  {Berti}} \emph {et~al.},\ }\href {\doibase 10.1088/0264-9381/32/24/243001}
  {\bibfield  {journal} {\bibinfo  {journal} {Class. Quant. Grav.}\ }\textbf
  {\bibinfo {volume} {32}},\ \bibinfo {pages} {243001} (\bibinfo {year}
  {2015})},\ \Eprint {http://arxiv.org/abs/1501.07274} {arXiv:1501.07274
  [gr-qc]} \BibitemShut {NoStop}%
\bibitem [{\citenamefont {Koyama}(2016)}]{Koyama:2015vza}%
  \BibitemOpen
  \bibfield  {author} {\bibinfo {author} {\bibfnamefont {K.}~\bibnamefont
  {Koyama}},\ }\href {\doibase 10.1088/0034-4885/79/4/046902} {\bibfield
  {journal} {\bibinfo  {journal} {Rept. Prog. Phys.}\ }\textbf {\bibinfo
  {volume} {79}},\ \bibinfo {pages} {046902} (\bibinfo {year} {2016})},\
  \Eprint {http://arxiv.org/abs/1504.04623} {arXiv:1504.04623 [astro-ph.CO]}
  \BibitemShut {NoStop}%
\bibitem [{\citenamefont {Lombriser}\ and\ \citenamefont
  {Taylor}(2016)}]{Lombriser:2015sxa}%
  \BibitemOpen
  \bibfield  {author} {\bibinfo {author} {\bibfnamefont {L.}~\bibnamefont
  {Lombriser}}\ and\ \bibinfo {author} {\bibfnamefont {A.}~\bibnamefont
  {Taylor}},\ }\href {\doibase 10.1088/1475-7516/2016/03/031} {\bibfield
  {journal} {\bibinfo  {journal} {JCAP}\ }\textbf {\bibinfo {volume} {1603}},\
  \bibinfo {pages} {031} (\bibinfo {year} {2016})},\ \Eprint
  {http://arxiv.org/abs/1509.08458} {arXiv:1509.08458 [astro-ph.CO]}
  \BibitemShut {NoStop}%
\bibitem [{\citenamefont {Lombriser}\ and\ \citenamefont
  {Lima}(2017)}]{Lombriser:2016yzn}%
  \BibitemOpen
  \bibfield  {author} {\bibinfo {author} {\bibfnamefont {L.}~\bibnamefont
  {Lombriser}}\ and\ \bibinfo {author} {\bibfnamefont {N.~A.}\ \bibnamefont
  {Lima}},\ }\href {\doibase 10.1016/j.physletb.2016.12.048} {\bibfield
  {journal} {\bibinfo  {journal} {Phys. Lett.}\ }\textbf {\bibinfo {volume}
  {B765}},\ \bibinfo {pages} {382} (\bibinfo {year} {2017})},\ \Eprint
  {http://arxiv.org/abs/1602.07670} {arXiv:1602.07670 [astro-ph.CO]}
  \BibitemShut {NoStop}%
\bibitem [{\citenamefont {Creminelli}\ and\ \citenamefont
  {Vernizzi}(2017)}]{Creminelli:2017sry}%
  \BibitemOpen
  \bibfield  {author} {\bibinfo {author} {\bibfnamefont {P.}~\bibnamefont
  {Creminelli}}\ and\ \bibinfo {author} {\bibfnamefont {F.}~\bibnamefont
  {Vernizzi}},\ }\href {\doibase 10.1103/PhysRevLett.119.251302} {\bibfield
  {journal} {\bibinfo  {journal} {Phys. Rev. Lett.}\ }\textbf {\bibinfo
  {volume} {119}},\ \bibinfo {pages} {251302} (\bibinfo {year} {2017})},\
  \Eprint {http://arxiv.org/abs/1710.05877} {arXiv:1710.05877 [astro-ph.CO]}
  \BibitemShut {NoStop}%
\bibitem [{\citenamefont {Sakstein}\ and\ \citenamefont
  {Jain}(2017)}]{Sakstein:2017xjx}%
  \BibitemOpen
  \bibfield  {author} {\bibinfo {author} {\bibfnamefont {J.}~\bibnamefont
  {Sakstein}}\ and\ \bibinfo {author} {\bibfnamefont {B.}~\bibnamefont
  {Jain}},\ }\href {\doibase 10.1103/PhysRevLett.119.251303} {\bibfield
  {journal} {\bibinfo  {journal} {Phys. Rev. Lett.}\ }\textbf {\bibinfo
  {volume} {119}},\ \bibinfo {pages} {251303} (\bibinfo {year} {2017})},\
  \Eprint {http://arxiv.org/abs/1710.05893} {arXiv:1710.05893 [astro-ph.CO]}
  \BibitemShut {NoStop}%
\bibitem [{\citenamefont {Ezquiaga}\ and\ \citenamefont
  {Zumalacárregui}(2017)}]{Ezquiaga:2017ekz}%
  \BibitemOpen
  \bibfield  {author} {\bibinfo {author} {\bibfnamefont {J.~M.}\ \bibnamefont
  {Ezquiaga}}\ and\ \bibinfo {author} {\bibfnamefont {M.}~\bibnamefont
  {Zumalacárregui}},\ }\href {\doibase 10.1103/PhysRevLett.119.251304}
  {\bibfield  {journal} {\bibinfo  {journal} {Phys. Rev. Lett.}\ }\textbf
  {\bibinfo {volume} {119}},\ \bibinfo {pages} {251304} (\bibinfo {year}
  {2017})},\ \Eprint {http://arxiv.org/abs/1710.05901} {arXiv:1710.05901
  [astro-ph.CO]} \BibitemShut {NoStop}%
\bibitem [{\citenamefont {Baker}\ \emph {et~al.}(2017)\citenamefont {Baker},
  \citenamefont {Bellini}, \citenamefont {Ferreira}, \citenamefont {Lagos},
  \citenamefont {Noller},\ and\ \citenamefont {Sawicki}}]{Baker:2017hug}%
  \BibitemOpen
  \bibfield  {author} {\bibinfo {author} {\bibfnamefont {T.}~\bibnamefont
  {Baker}}, \bibinfo {author} {\bibfnamefont {E.}~\bibnamefont {Bellini}},
  \bibinfo {author} {\bibfnamefont {P.~G.}\ \bibnamefont {Ferreira}}, \bibinfo
  {author} {\bibfnamefont {M.}~\bibnamefont {Lagos}}, \bibinfo {author}
  {\bibfnamefont {J.}~\bibnamefont {Noller}}, \ and\ \bibinfo {author}
  {\bibfnamefont {I.}~\bibnamefont {Sawicki}},\ }\href {\doibase
  10.1103/PhysRevLett.119.251301} {\bibfield  {journal} {\bibinfo  {journal}
  {Phys. Rev. Lett.}\ }\textbf {\bibinfo {volume} {119}},\ \bibinfo {pages}
  {251301} (\bibinfo {year} {2017})},\ \Eprint
  {http://arxiv.org/abs/1710.06394} {arXiv:1710.06394 [astro-ph.CO]}
  \BibitemShut {NoStop}%
\bibitem [{\citenamefont {Crisostomi}\ and\ \citenamefont
  {Koyama}(2018)}]{Crisostomi:2017lbg}%
  \BibitemOpen
  \bibfield  {author} {\bibinfo {author} {\bibfnamefont {M.}~\bibnamefont
  {Crisostomi}}\ and\ \bibinfo {author} {\bibfnamefont {K.}~\bibnamefont
  {Koyama}},\ }\href {\doibase 10.1103/PhysRevD.97.021301} {\bibfield
  {journal} {\bibinfo  {journal} {Phys. Rev.}\ }\textbf {\bibinfo {volume}
  {D97}},\ \bibinfo {pages} {021301} (\bibinfo {year} {2018})},\ \Eprint
  {http://arxiv.org/abs/1711.06661} {arXiv:1711.06661 [astro-ph.CO]}
  \BibitemShut {NoStop}%
\bibitem [{\citenamefont {Langlois}\ \emph {et~al.}(2018)\citenamefont
  {Langlois}, \citenamefont {Saito}, \citenamefont {Yamauchi},\ and\
  \citenamefont {Noui}}]{Langlois:2017dyl}%
  \BibitemOpen
  \bibfield  {author} {\bibinfo {author} {\bibfnamefont {D.}~\bibnamefont
  {Langlois}}, \bibinfo {author} {\bibfnamefont {R.}~\bibnamefont {Saito}},
  \bibinfo {author} {\bibfnamefont {D.}~\bibnamefont {Yamauchi}}, \ and\
  \bibinfo {author} {\bibfnamefont {K.}~\bibnamefont {Noui}},\ }\href {\doibase
  10.1103/PhysRevD.97.061501} {\bibfield  {journal} {\bibinfo  {journal} {Phys.
  Rev.}\ }\textbf {\bibinfo {volume} {D97}},\ \bibinfo {pages} {061501}
  (\bibinfo {year} {2018})},\ \Eprint {http://arxiv.org/abs/1711.07403}
  {arXiv:1711.07403 [gr-qc]} \BibitemShut {NoStop}%
\bibitem [{\citenamefont {Bartolo}\ \emph {et~al.}(2018)\citenamefont
  {Bartolo}, \citenamefont {Karmakar}, \citenamefont {Matarrese},\ and\
  \citenamefont {Scomparin}}]{Bartolo:2017ibw}%
  \BibitemOpen
  \bibfield  {author} {\bibinfo {author} {\bibfnamefont {N.}~\bibnamefont
  {Bartolo}}, \bibinfo {author} {\bibfnamefont {P.}~\bibnamefont {Karmakar}},
  \bibinfo {author} {\bibfnamefont {S.}~\bibnamefont {Matarrese}}, \ and\
  \bibinfo {author} {\bibfnamefont {M.}~\bibnamefont {Scomparin}},\ }\href
  {\doibase 10.1088/1475-7516/2018/05/048} {\bibfield  {journal} {\bibinfo
  {journal} {JCAP}\ }\textbf {\bibinfo {volume} {1805}},\ \bibinfo {pages}
  {048} (\bibinfo {year} {2018})},\ \Eprint {http://arxiv.org/abs/1712.04002}
  {arXiv:1712.04002 [gr-qc]} \BibitemShut {NoStop}%
\bibitem [{\citenamefont {Dima}\ and\ \citenamefont
  {Vernizzi}(2018)}]{Dima:2017pwp}%
  \BibitemOpen
  \bibfield  {author} {\bibinfo {author} {\bibfnamefont {A.}~\bibnamefont
  {Dima}}\ and\ \bibinfo {author} {\bibfnamefont {F.}~\bibnamefont
  {Vernizzi}},\ }\href {\doibase 10.1103/PhysRevD.97.101302} {\bibfield
  {journal} {\bibinfo  {journal} {Phys. Rev.}\ }\textbf {\bibinfo {volume}
  {D97}},\ \bibinfo {pages} {101302} (\bibinfo {year} {2018})},\ \Eprint
  {http://arxiv.org/abs/1712.04731} {arXiv:1712.04731 [gr-qc]} \BibitemShut
  {NoStop}%
\bibitem [{\citenamefont {Aso}\ \emph {et~al.}(2013)\citenamefont {Aso},
  \citenamefont {Michimura}, \citenamefont {Somiya}, \citenamefont {Ando},
  \citenamefont {Miyakawa}, \citenamefont {Sekiguchi}, \citenamefont
  {Tatsumi},\ and\ \citenamefont {Yamamoto}}]{Aso:2013eba}%
  \BibitemOpen
  \bibfield  {author} {\bibinfo {author} {\bibfnamefont {Y.}~\bibnamefont
  {Aso}}, \bibinfo {author} {\bibfnamefont {Y.}~\bibnamefont {Michimura}},
  \bibinfo {author} {\bibfnamefont {K.}~\bibnamefont {Somiya}}, \bibinfo
  {author} {\bibfnamefont {M.}~\bibnamefont {Ando}}, \bibinfo {author}
  {\bibfnamefont {O.}~\bibnamefont {Miyakawa}}, \bibinfo {author}
  {\bibfnamefont {T.}~\bibnamefont {Sekiguchi}}, \bibinfo {author}
  {\bibfnamefont {D.}~\bibnamefont {Tatsumi}}, \ and\ \bibinfo {author}
  {\bibfnamefont {H.}~\bibnamefont {Yamamoto}} (\bibinfo {collaboration}
  {KAGRA}),\ }\href {\doibase 10.1103/PhysRevD.88.043007} {\bibfield  {journal}
  {\bibinfo  {journal} {Phys. Rev.}\ }\textbf {\bibinfo {volume} {D88}},\
  \bibinfo {pages} {043007} (\bibinfo {year} {2013})},\ \Eprint
  {http://arxiv.org/abs/1306.6747} {arXiv:1306.6747 [gr-qc]} \BibitemShut
  {NoStop}%
\bibitem [{\citenamefont {Barausse}\ and\ \citenamefont
  {Sotiriou}(2008)}]{Barausse:2008xv}%
  \BibitemOpen
  \bibfield  {author} {\bibinfo {author} {\bibfnamefont {E.}~\bibnamefont
  {Barausse}}\ and\ \bibinfo {author} {\bibfnamefont {T.~P.}\ \bibnamefont
  {Sotiriou}},\ }\href {\doibase 10.1103/PhysRevLett.101.099001} {\bibfield
  {journal} {\bibinfo  {journal} {Phys. Rev. Lett.}\ }\textbf {\bibinfo
  {volume} {101}},\ \bibinfo {pages} {099001} (\bibinfo {year} {2008})},\
  \Eprint {http://arxiv.org/abs/0803.3433} {arXiv:0803.3433 [gr-qc]}
  \BibitemShut {NoStop}%
\bibitem [{\citenamefont {Israel}(1967)}]{Israel:1967wq}%
  \BibitemOpen
  \bibfield  {author} {\bibinfo {author} {\bibfnamefont {W.}~\bibnamefont
  {Israel}},\ }\href {\doibase 10.1103/PhysRev.164.1776} {\bibfield  {journal}
  {\bibinfo  {journal} {Phys. Rev.}\ }\textbf {\bibinfo {volume} {164}},\
  \bibinfo {pages} {1776} (\bibinfo {year} {1967})}\BibitemShut {NoStop}%
\bibitem [{\citenamefont {Carter}(1971)}]{Carter:1971zc}%
  \BibitemOpen
  \bibfield  {author} {\bibinfo {author} {\bibfnamefont {B.}~\bibnamefont
  {Carter}},\ }\href {\doibase 10.1103/PhysRevLett.26.331} {\bibfield
  {journal} {\bibinfo  {journal} {Phys. Rev. Lett.}\ }\textbf {\bibinfo
  {volume} {26}},\ \bibinfo {pages} {331} (\bibinfo {year} {1971})}\BibitemShut
  {NoStop}%
\bibitem [{\citenamefont {Hawking}(1972{\natexlab{a}})}]{Hawking:1971vc}%
  \BibitemOpen
  \bibfield  {author} {\bibinfo {author} {\bibfnamefont {S.~W.}\ \bibnamefont
  {Hawking}},\ }\href {\doibase 10.1007/BF01877517} {\bibfield  {journal}
  {\bibinfo  {journal} {Commun. Math. Phys.}\ }\textbf {\bibinfo {volume}
  {25}},\ \bibinfo {pages} {152} (\bibinfo {year}
  {1972}{\natexlab{a}})}\BibitemShut {NoStop}%
\bibitem [{\citenamefont {Bocharova}\ \emph {et~al.}(1970)\citenamefont
  {Bocharova}, \citenamefont {Bronnikov},\ and\ \citenamefont
  {Melnikov}}]{Bocharova:1970skc}%
  \BibitemOpen
  \bibfield  {author} {\bibinfo {author} {\bibfnamefont {N.~M.}\ \bibnamefont
  {Bocharova}}, \bibinfo {author} {\bibfnamefont {K.~A.}\ \bibnamefont
  {Bronnikov}}, \ and\ \bibinfo {author} {\bibfnamefont {V.~N.}\ \bibnamefont
  {Melnikov}},\ }\href@noop {} {\bibfield  {journal} {\bibinfo  {journal}
  {Vestn. Mosk. Univ. Ser. III Fiz. Astron.}\ ,\ \bibinfo {pages} {706}}
  (\bibinfo {year} {1970})}\BibitemShut {NoStop}%
\bibitem [{\citenamefont {Bekenstein}(1974)}]{Bekenstein:1974sf}%
  \BibitemOpen
  \bibfield  {author} {\bibinfo {author} {\bibfnamefont {J.~D.}\ \bibnamefont
  {Bekenstein}},\ }\href {\doibase 10.1016/0003-4916(74)90124-9} {\bibfield
  {journal} {\bibinfo  {journal} {Annals Phys.}\ }\textbf {\bibinfo {volume}
  {82}},\ \bibinfo {pages} {535} (\bibinfo {year} {1974})}\BibitemShut
  {NoStop}%
\bibitem [{\citenamefont {Kanti}\ \emph {et~al.}(1996)\citenamefont {Kanti},
  \citenamefont {Mavromatos}, \citenamefont {Rizos}, \citenamefont {Tamvakis},\
  and\ \citenamefont {Winstanley}}]{Kanti:1995vq}%
  \BibitemOpen
  \bibfield  {author} {\bibinfo {author} {\bibfnamefont {P.}~\bibnamefont
  {Kanti}}, \bibinfo {author} {\bibfnamefont {N.~E.}\ \bibnamefont
  {Mavromatos}}, \bibinfo {author} {\bibfnamefont {J.}~\bibnamefont {Rizos}},
  \bibinfo {author} {\bibfnamefont {K.}~\bibnamefont {Tamvakis}}, \ and\
  \bibinfo {author} {\bibfnamefont {E.}~\bibnamefont {Winstanley}},\ }\href
  {\doibase 10.1103/PhysRevD.54.5049} {\bibfield  {journal} {\bibinfo
  {journal} {Phys. Rev.}\ }\textbf {\bibinfo {volume} {D54}},\ \bibinfo {pages}
  {5049} (\bibinfo {year} {1996})},\ \Eprint
  {http://arxiv.org/abs/hep-th/9511071} {arXiv:hep-th/9511071 [hep-th]}
  \BibitemShut {NoStop}%
\bibitem [{\citenamefont {Pani}\ and\ \citenamefont
  {Cardoso}(2009)}]{Pani:2009wy}%
  \BibitemOpen
  \bibfield  {author} {\bibinfo {author} {\bibfnamefont {P.}~\bibnamefont
  {Pani}}\ and\ \bibinfo {author} {\bibfnamefont {V.}~\bibnamefont {Cardoso}},\
  }\href {\doibase 10.1103/PhysRevD.79.084031} {\bibfield  {journal} {\bibinfo
  {journal} {Phys. Rev.}\ }\textbf {\bibinfo {volume} {D79}},\ \bibinfo {pages}
  {084031} (\bibinfo {year} {2009})},\ \Eprint {http://arxiv.org/abs/0902.1569}
  {arXiv:0902.1569 [gr-qc]} \BibitemShut {NoStop}%
\bibitem [{\citenamefont {Kleihaus}\ \emph {et~al.}(2011)\citenamefont
  {Kleihaus}, \citenamefont {Kunz},\ and\ \citenamefont
  {Radu}}]{Kleihaus:2011tg}%
  \BibitemOpen
  \bibfield  {author} {\bibinfo {author} {\bibfnamefont {B.}~\bibnamefont
  {Kleihaus}}, \bibinfo {author} {\bibfnamefont {J.}~\bibnamefont {Kunz}}, \
  and\ \bibinfo {author} {\bibfnamefont {E.}~\bibnamefont {Radu}},\ }\href
  {\doibase 10.1103/PhysRevLett.106.151104} {\bibfield  {journal} {\bibinfo
  {journal} {Phys. Rev. Lett.}\ }\textbf {\bibinfo {volume} {106}},\ \bibinfo
  {pages} {151104} (\bibinfo {year} {2011})},\ \Eprint
  {http://arxiv.org/abs/1101.2868} {arXiv:1101.2868 [gr-qc]} \BibitemShut
  {NoStop}%
\bibitem [{\citenamefont {Ayzenberg}\ and\ \citenamefont
  {Yunes}(2014)}]{Ayzenberg:2014aka}%
  \BibitemOpen
  \bibfield  {author} {\bibinfo {author} {\bibfnamefont {D.}~\bibnamefont
  {Ayzenberg}}\ and\ \bibinfo {author} {\bibfnamefont {N.}~\bibnamefont
  {Yunes}},\ }\href {\doibase 10.1103/PhysRevD.91.069905,
  10.1103/PhysRevD.90.044066} {\bibfield  {journal} {\bibinfo  {journal} {Phys.
  Rev.}\ }\textbf {\bibinfo {volume} {D90}},\ \bibinfo {pages} {044066}
  (\bibinfo {year} {2014})},\ \bibinfo {note} {[Erratum: Phys.
  Rev.D91,no.6,069905(2015)]},\ \Eprint {http://arxiv.org/abs/1405.2133}
  {arXiv:1405.2133 [gr-qc]} \BibitemShut {NoStop}%
\bibitem [{\citenamefont {Sotiriou}\ and\ \citenamefont
  {Zhou}(2014)}]{Sotiriou:2013qea}%
  \BibitemOpen
  \bibfield  {author} {\bibinfo {author} {\bibfnamefont {T.~P.}\ \bibnamefont
  {Sotiriou}}\ and\ \bibinfo {author} {\bibfnamefont {S.-Y.}\ \bibnamefont
  {Zhou}},\ }\href {\doibase 10.1103/PhysRevLett.112.251102} {\bibfield
  {journal} {\bibinfo  {journal} {Phys. Rev. Lett.}\ }\textbf {\bibinfo
  {volume} {112}},\ \bibinfo {pages} {251102} (\bibinfo {year} {2014})},\
  \Eprint {http://arxiv.org/abs/1312.3622} {arXiv:1312.3622 [gr-qc]}
  \BibitemShut {NoStop}%
\bibitem [{\citenamefont {Anabalon}\ \emph {et~al.}(2014)\citenamefont
  {Anabalon}, \citenamefont {Cisterna},\ and\ \citenamefont
  {Oliva}}]{Anabalon:2013oea}%
  \BibitemOpen
  \bibfield  {author} {\bibinfo {author} {\bibfnamefont {A.}~\bibnamefont
  {Anabalon}}, \bibinfo {author} {\bibfnamefont {A.}~\bibnamefont {Cisterna}},
  \ and\ \bibinfo {author} {\bibfnamefont {J.}~\bibnamefont {Oliva}},\ }\href
  {\doibase 10.1103/PhysRevD.89.084050} {\bibfield  {journal} {\bibinfo
  {journal} {Phys. Rev.}\ }\textbf {\bibinfo {volume} {D89}},\ \bibinfo {pages}
  {084050} (\bibinfo {year} {2014})},\ \Eprint {http://arxiv.org/abs/1312.3597}
  {arXiv:1312.3597 [gr-qc]} \BibitemShut {NoStop}%
\bibitem [{\citenamefont {Minamitsuji}(2014)}]{Minamitsuji:2013ura}%
  \BibitemOpen
  \bibfield  {author} {\bibinfo {author} {\bibfnamefont {M.}~\bibnamefont
  {Minamitsuji}},\ }\href {\doibase 10.1103/PhysRevD.89.064017} {\bibfield
  {journal} {\bibinfo  {journal} {Phys. Rev.}\ }\textbf {\bibinfo {volume}
  {D89}},\ \bibinfo {pages} {064017} (\bibinfo {year} {2014})},\ \Eprint
  {http://arxiv.org/abs/1312.3759} {arXiv:1312.3759 [gr-qc]} \BibitemShut
  {NoStop}%
\bibitem [{\citenamefont {Erices}\ and\ \citenamefont
  {Martinez}(2018)}]{Erices:2017izj}%
  \BibitemOpen
  \bibfield  {author} {\bibinfo {author} {\bibfnamefont {C.}~\bibnamefont
  {Erices}}\ and\ \bibinfo {author} {\bibfnamefont {C.}~\bibnamefont
  {Martinez}},\ }\href {\doibase 10.1103/PhysRevD.97.024034} {\bibfield
  {journal} {\bibinfo  {journal} {Phys. Rev.}\ }\textbf {\bibinfo {volume}
  {D97}},\ \bibinfo {pages} {024034} (\bibinfo {year} {2018})},\ \Eprint
  {http://arxiv.org/abs/1707.03483} {arXiv:1707.03483 [hep-th]} \BibitemShut
  {NoStop}%
\bibitem [{\citenamefont {Babichev}\ and\ \citenamefont
  {Charmousis}(2014)}]{Babichev:2013cya}%
  \BibitemOpen
  \bibfield  {author} {\bibinfo {author} {\bibfnamefont {E.}~\bibnamefont
  {Babichev}}\ and\ \bibinfo {author} {\bibfnamefont {C.}~\bibnamefont
  {Charmousis}},\ }\href {\doibase 10.1007/JHEP08(2014)106} {\bibfield
  {journal} {\bibinfo  {journal} {JHEP}\ }\textbf {\bibinfo {volume} {08}},\
  \bibinfo {pages} {106} (\bibinfo {year} {2014})},\ \Eprint
  {http://arxiv.org/abs/1312.3204} {arXiv:1312.3204 [gr-qc]} \BibitemShut
  {NoStop}%
\bibitem [{\citenamefont {Herdeiro}\ and\ \citenamefont
  {Radu}(2014)}]{Herdeiro:2014goa}%
  \BibitemOpen
  \bibfield  {author} {\bibinfo {author} {\bibfnamefont {C.~A.~R.}\
  \bibnamefont {Herdeiro}}\ and\ \bibinfo {author} {\bibfnamefont
  {E.}~\bibnamefont {Radu}},\ }\href {\doibase 10.1103/PhysRevLett.112.221101}
  {\bibfield  {journal} {\bibinfo  {journal} {Phys. Rev. Lett.}\ }\textbf
  {\bibinfo {volume} {112}},\ \bibinfo {pages} {221101} (\bibinfo {year}
  {2014})},\ \Eprint {http://arxiv.org/abs/1403.2757} {arXiv:1403.2757 [gr-qc]}
  \BibitemShut {NoStop}%
\bibitem [{\citenamefont {Charmousis}\ \emph {et~al.}(2014)\citenamefont
  {Charmousis}, \citenamefont {Kolyvaris}, \citenamefont {Papantonopoulos},\
  and\ \citenamefont {Tsoukalas}}]{Charmousis:2014zaa}%
  \BibitemOpen
  \bibfield  {author} {\bibinfo {author} {\bibfnamefont {C.}~\bibnamefont
  {Charmousis}}, \bibinfo {author} {\bibfnamefont {T.}~\bibnamefont
  {Kolyvaris}}, \bibinfo {author} {\bibfnamefont {E.}~\bibnamefont
  {Papantonopoulos}}, \ and\ \bibinfo {author} {\bibfnamefont {M.}~\bibnamefont
  {Tsoukalas}},\ }\href {\doibase 10.1007/JHEP07(2014)085} {\bibfield
  {journal} {\bibinfo  {journal} {JHEP}\ }\textbf {\bibinfo {volume} {07}},\
  \bibinfo {pages} {085} (\bibinfo {year} {2014})},\ \Eprint
  {http://arxiv.org/abs/1404.1024} {arXiv:1404.1024 [gr-qc]} \BibitemShut
  {NoStop}%
\bibitem [{\citenamefont {Radu}\ and\ \citenamefont
  {Winstanley}(2005)}]{Radu:2005bp}%
  \BibitemOpen
  \bibfield  {author} {\bibinfo {author} {\bibfnamefont {E.}~\bibnamefont
  {Radu}}\ and\ \bibinfo {author} {\bibfnamefont {E.}~\bibnamefont
  {Winstanley}},\ }\href {\doibase 10.1103/PhysRevD.72.024017} {\bibfield
  {journal} {\bibinfo  {journal} {Phys. Rev.}\ }\textbf {\bibinfo {volume}
  {D72}},\ \bibinfo {pages} {024017} (\bibinfo {year} {2005})},\ \Eprint
  {http://arxiv.org/abs/gr-qc/0503095} {arXiv:gr-qc/0503095 [gr-qc]}
  \BibitemShut {NoStop}%
\bibitem [{\citenamefont {Anabalon}\ and\ \citenamefont
  {Maeda}(2010)}]{Anabalon:2009qt}%
  \BibitemOpen
  \bibfield  {author} {\bibinfo {author} {\bibfnamefont {A.}~\bibnamefont
  {Anabalon}}\ and\ \bibinfo {author} {\bibfnamefont {H.}~\bibnamefont
  {Maeda}},\ }\href {\doibase 10.1103/PhysRevD.81.041501} {\bibfield  {journal}
  {\bibinfo  {journal} {Phys. Rev.}\ }\textbf {\bibinfo {volume} {D81}},\
  \bibinfo {pages} {041501} (\bibinfo {year} {2010})},\ \Eprint
  {http://arxiv.org/abs/0907.0219} {arXiv:0907.0219 [hep-th]} \BibitemShut
  {NoStop}%
\bibitem [{\citenamefont {Kolyvaris}\ \emph {et~al.}(2012)\citenamefont
  {Kolyvaris}, \citenamefont {Koutsoumbas}, \citenamefont {Papantonopoulos},\
  and\ \citenamefont {Siopsis}}]{Kolyvaris:2011fk}%
  \BibitemOpen
  \bibfield  {author} {\bibinfo {author} {\bibfnamefont {T.}~\bibnamefont
  {Kolyvaris}}, \bibinfo {author} {\bibfnamefont {G.}~\bibnamefont
  {Koutsoumbas}}, \bibinfo {author} {\bibfnamefont {E.}~\bibnamefont
  {Papantonopoulos}}, \ and\ \bibinfo {author} {\bibfnamefont {G.}~\bibnamefont
  {Siopsis}},\ }\href {\doibase 10.1088/0264-9381/29/20/205011} {\bibfield
  {journal} {\bibinfo  {journal} {Class. Quant. Grav.}\ }\textbf {\bibinfo
  {volume} {29}},\ \bibinfo {pages} {205011} (\bibinfo {year} {2012})},\
  \Eprint {http://arxiv.org/abs/1111.0263} {arXiv:1111.0263 [gr-qc]}
  \BibitemShut {NoStop}%
\bibitem [{\citenamefont {Anabalon}(2012)}]{Anabalon:2012ta}%
  \BibitemOpen
  \bibfield  {author} {\bibinfo {author} {\bibfnamefont {A.}~\bibnamefont
  {Anabalon}},\ }\href {\doibase 10.1007/JHEP06(2012)127} {\bibfield  {journal}
  {\bibinfo  {journal} {JHEP}\ }\textbf {\bibinfo {volume} {06}},\ \bibinfo
  {pages} {127} (\bibinfo {year} {2012})},\ \Eprint
  {http://arxiv.org/abs/1204.2720} {arXiv:1204.2720 [hep-th]} \BibitemShut
  {NoStop}%
\bibitem [{\citenamefont {Babichev}\ \emph {et~al.}(2016)\citenamefont
  {Babichev}, \citenamefont {Charmousis}, \citenamefont {Lehebel},\ and\
  \citenamefont {Moskalets}}]{Babichev:2016fbg}%
  \BibitemOpen
  \bibfield  {author} {\bibinfo {author} {\bibfnamefont {E.}~\bibnamefont
  {Babichev}}, \bibinfo {author} {\bibfnamefont {C.}~\bibnamefont
  {Charmousis}}, \bibinfo {author} {\bibfnamefont {A.}~\bibnamefont {Lehebel}},
  \ and\ \bibinfo {author} {\bibfnamefont {T.}~\bibnamefont {Moskalets}},\
  }\href {\doibase 10.1088/1475-7516/2016/09/011} {\bibfield  {journal}
  {\bibinfo  {journal} {JCAP}\ }\textbf {\bibinfo {volume} {1609}},\ \bibinfo
  {pages} {011} (\bibinfo {year} {2016})},\ \Eprint
  {http://arxiv.org/abs/1605.07438} {arXiv:1605.07438 [gr-qc]} \BibitemShut
  {NoStop}%
\bibitem [{\citenamefont {Babichev}\ \emph {et~al.}(2017)\citenamefont
  {Babichev}, \citenamefont {Charmousis},\ and\ \citenamefont
  {Lehebel}}]{Babichev:2017guv}%
  \BibitemOpen
  \bibfield  {author} {\bibinfo {author} {\bibfnamefont {E.}~\bibnamefont
  {Babichev}}, \bibinfo {author} {\bibfnamefont {C.}~\bibnamefont
  {Charmousis}}, \ and\ \bibinfo {author} {\bibfnamefont {A.}~\bibnamefont
  {Lehebel}},\ }\href {\doibase 10.1088/1475-7516/2017/04/027} {\bibfield
  {journal} {\bibinfo  {journal} {JCAP}\ }\textbf {\bibinfo {volume} {1704}},\
  \bibinfo {pages} {027} (\bibinfo {year} {2017})},\ \Eprint
  {http://arxiv.org/abs/1702.01938} {arXiv:1702.01938 [gr-qc]} \BibitemShut
  {NoStop}%
\bibitem [{\citenamefont {Chase}(1970)}]{Chase}%
  \BibitemOpen
  \bibfield  {author} {\bibinfo {author} {\bibfnamefont {J.~E.}\ \bibnamefont
  {Chase}},\ }\href {\doibase 10.1007/BF01646635} {\bibfield  {journal}
  {\bibinfo  {journal} {Commun. Math. Phys.}\ }\textbf {\bibinfo {volume}
  {22}},\ \bibinfo {pages} {276} (\bibinfo {year} {1970})}\BibitemShut
  {NoStop}%
\bibitem [{\citenamefont {Bekenstein}(1972)}]{Bekenstein:1972ny}%
  \BibitemOpen
  \bibfield  {author} {\bibinfo {author} {\bibfnamefont {J.~D.}\ \bibnamefont
  {Bekenstein}},\ }\href {\doibase 10.1103/PhysRevLett.28.452} {\bibfield
  {journal} {\bibinfo  {journal} {Phys. Rev. Lett.}\ }\textbf {\bibinfo
  {volume} {28}},\ \bibinfo {pages} {452} (\bibinfo {year} {1972})}\BibitemShut
  {NoStop}%
\bibitem [{\citenamefont {Graham}\ and\ \citenamefont
  {Jha}(2014)}]{Graham:2014mda}%
  \BibitemOpen
  \bibfield  {author} {\bibinfo {author} {\bibfnamefont {A.~A.~H.}\
  \bibnamefont {Graham}}\ and\ \bibinfo {author} {\bibfnamefont
  {R.}~\bibnamefont {Jha}},\ }\href {\doibase 10.1103/PhysRevD.92.069901,
  10.1103/PhysRevD.89.084056} {\bibfield  {journal} {\bibinfo  {journal} {Phys.
  Rev.}\ }\textbf {\bibinfo {volume} {D89}},\ \bibinfo {pages} {084056}
  (\bibinfo {year} {2014})},\ \bibinfo {note} {[Erratum: Phys.
  Rev.D92,no.6,069901(2015)]},\ \Eprint {http://arxiv.org/abs/1401.8203}
  {arXiv:1401.8203 [gr-qc]} \BibitemShut {NoStop}%
\bibitem [{\citenamefont {Hawking}(1972{\natexlab{b}})}]{Hawking:1972qk}%
  \BibitemOpen
  \bibfield  {author} {\bibinfo {author} {\bibfnamefont {S.~W.}\ \bibnamefont
  {Hawking}},\ }\href {\doibase 10.1007/BF01877518} {\bibfield  {journal}
  {\bibinfo  {journal} {Commun. Math. Phys.}\ }\textbf {\bibinfo {volume}
  {25}},\ \bibinfo {pages} {167} (\bibinfo {year}
  {1972}{\natexlab{b}})}\BibitemShut {NoStop}%
\bibitem [{\citenamefont {Bekenstein}(1995)}]{Bekenstein:1995un}%
  \BibitemOpen
  \bibfield  {author} {\bibinfo {author} {\bibfnamefont {J.~D.}\ \bibnamefont
  {Bekenstein}},\ }\href {\doibase 10.1103/PhysRevD.51.R6608} {\bibfield
  {journal} {\bibinfo  {journal} {Phys. Rev.}\ }\textbf {\bibinfo {volume}
  {D51}},\ \bibinfo {pages} {R6608} (\bibinfo {year} {1995})}\BibitemShut
  {NoStop}%
\bibitem [{\citenamefont {Sotiriou}\ and\ \citenamefont
  {Faraoni}(2012)}]{Sotiriou:2011dz}%
  \BibitemOpen
  \bibfield  {author} {\bibinfo {author} {\bibfnamefont {T.~P.}\ \bibnamefont
  {Sotiriou}}\ and\ \bibinfo {author} {\bibfnamefont {V.}~\bibnamefont
  {Faraoni}},\ }\href {\doibase 10.1103/PhysRevLett.108.081103} {\bibfield
  {journal} {\bibinfo  {journal} {Phys. Rev. Lett.}\ }\textbf {\bibinfo
  {volume} {108}},\ \bibinfo {pages} {081103} (\bibinfo {year} {2012})},\
  \Eprint {http://arxiv.org/abs/1109.6324} {arXiv:1109.6324 [gr-qc]}
  \BibitemShut {NoStop}%
\bibitem [{\citenamefont {Faraoni}(2017)}]{Faraoni:2017ock}%
  \BibitemOpen
  \bibfield  {author} {\bibinfo {author} {\bibfnamefont {V.}~\bibnamefont
  {Faraoni}},\ }\href {\doibase 10.1103/PhysRevD.95.124013} {\bibfield
  {journal} {\bibinfo  {journal} {Phys. Rev.}\ }\textbf {\bibinfo {volume}
  {D95}},\ \bibinfo {pages} {124013} (\bibinfo {year} {2017})},\ \Eprint
  {http://arxiv.org/abs/1705.07134} {arXiv:1705.07134 [gr-qc]} \BibitemShut
  {NoStop}%
\bibitem [{\citenamefont {Hui}\ and\ \citenamefont
  {Nicolis}(2013)}]{Hui:2012qt}%
  \BibitemOpen
  \bibfield  {author} {\bibinfo {author} {\bibfnamefont {L.}~\bibnamefont
  {Hui}}\ and\ \bibinfo {author} {\bibfnamefont {A.}~\bibnamefont {Nicolis}},\
  }\href {\doibase 10.1103/PhysRevLett.110.241104} {\bibfield  {journal}
  {\bibinfo  {journal} {Phys. Rev. Lett.}\ }\textbf {\bibinfo {volume} {110}},\
  \bibinfo {pages} {241104} (\bibinfo {year} {2013})},\ \Eprint
  {http://arxiv.org/abs/1202.1296} {arXiv:1202.1296 [hep-th]} \BibitemShut
  {NoStop}%
\bibitem [{\citenamefont {Herdeiro}\ and\ \citenamefont
  {Radu}(2015)}]{Herdeiro:2015waa}%
  \BibitemOpen
  \bibfield  {author} {\bibinfo {author} {\bibfnamefont {C.~A.~R.}\
  \bibnamefont {Herdeiro}}\ and\ \bibinfo {author} {\bibfnamefont
  {E.}~\bibnamefont {Radu}},\ }\bibfield  {booktitle} {\emph {\bibinfo
  {booktitle} {{Proceedings, 7th Black Holes Workshop 2014: Aveiro, Portugal,
  December 18-19, 2014}}},\ }\href {\doibase 10.1142/S0218271815420146}
  {\bibfield  {journal} {\bibinfo  {journal} {Int. J. Mod. Phys.}\ }\textbf
  {\bibinfo {volume} {D24}},\ \bibinfo {pages} {1542014} (\bibinfo {year}
  {2015})},\ \Eprint {http://arxiv.org/abs/1504.08209} {arXiv:1504.08209
  [gr-qc]} \BibitemShut {NoStop}%
\bibitem [{\citenamefont {Tattersall}\ \emph {et~al.}(2018)\citenamefont
  {Tattersall}, \citenamefont {Ferreira},\ and\ \citenamefont
  {Lagos}}]{Tattersall:2018map}%
  \BibitemOpen
  \bibfield  {author} {\bibinfo {author} {\bibfnamefont {O.~J.}\ \bibnamefont
  {Tattersall}}, \bibinfo {author} {\bibfnamefont {P.~G.}\ \bibnamefont
  {Ferreira}}, \ and\ \bibinfo {author} {\bibfnamefont {M.}~\bibnamefont
  {Lagos}},\ }\href {\doibase 10.1103/PhysRevD.97.084005} {\bibfield  {journal}
  {\bibinfo  {journal} {Phys. Rev.}\ }\textbf {\bibinfo {volume} {D97}},\
  \bibinfo {pages} {084005} (\bibinfo {year} {2018})},\ \Eprint
  {http://arxiv.org/abs/1802.08606} {arXiv:1802.08606 [gr-qc]} \BibitemShut
  {NoStop}%
\bibitem [{\citenamefont {Psaltis}\ \emph {et~al.}(2008)\citenamefont
  {Psaltis}, \citenamefont {Perrodin}, \citenamefont {Dienes},\ and\
  \citenamefont {Mocioiu}}]{Psaltis:2007cw}%
  \BibitemOpen
  \bibfield  {author} {\bibinfo {author} {\bibfnamefont {D.}~\bibnamefont
  {Psaltis}}, \bibinfo {author} {\bibfnamefont {D.}~\bibnamefont {Perrodin}},
  \bibinfo {author} {\bibfnamefont {K.~R.}\ \bibnamefont {Dienes}}, \ and\
  \bibinfo {author} {\bibfnamefont {I.}~\bibnamefont {Mocioiu}},\ }\href
  {\doibase 10.1103/PhysRevLett.100.091101, 10.1103/PhysRevLett.100.119902}
  {\bibfield  {journal} {\bibinfo  {journal} {Phys. Rev. Lett.}\ }\textbf
  {\bibinfo {volume} {100}},\ \bibinfo {pages} {091101} (\bibinfo {year}
  {2008})},\ \bibinfo {note} {[Phys. Rev. Lett.100,119902(2008)]},\ \Eprint
  {http://arxiv.org/abs/0710.4564} {arXiv:0710.4564 [astro-ph]} \BibitemShut
  {NoStop}%
\bibitem [{\citenamefont {Li}\ \emph {et~al.}(2018)\citenamefont {Li},
  \citenamefont {Liu},\ and\ \citenamefont {Lu}}]{Li:2017ncu}%
  \BibitemOpen
  \bibfield  {author} {\bibinfo {author} {\bibfnamefont {Y.-Z.}\ \bibnamefont
  {Li}}, \bibinfo {author} {\bibfnamefont {H.-S.}\ \bibnamefont {Liu}}, \ and\
  \bibinfo {author} {\bibfnamefont {H.}~\bibnamefont {Lu}},\ }\href {\doibase
  10.1007/JHEP02(2018)166} {\bibfield  {journal} {\bibinfo  {journal} {JHEP}\
  }\textbf {\bibinfo {volume} {02}},\ \bibinfo {pages} {166} (\bibinfo {year}
  {2018})},\ \Eprint {http://arxiv.org/abs/1708.07198} {arXiv:1708.07198
  [hep-th]} \BibitemShut {NoStop}%
\bibitem [{\citenamefont {Emparan}\ and\ \citenamefont
  {Reall}(2002)}]{Emparan:2001wn}%
  \BibitemOpen
  \bibfield  {author} {\bibinfo {author} {\bibfnamefont {R.}~\bibnamefont
  {Emparan}}\ and\ \bibinfo {author} {\bibfnamefont {H.~S.}\ \bibnamefont
  {Reall}},\ }\href {\doibase 10.1103/PhysRevLett.88.101101} {\bibfield
  {journal} {\bibinfo  {journal} {Phys. Rev. Lett.}\ }\textbf {\bibinfo
  {volume} {88}},\ \bibinfo {pages} {101101} (\bibinfo {year} {2002})},\
  \Eprint {http://arxiv.org/abs/hep-th/0110260} {arXiv:hep-th/0110260 [hep-th]}
  \BibitemShut {NoStop}%
\bibitem [{\citenamefont {Emparan}\ and\ \citenamefont
  {Reall}(2008)}]{Emparan:2008eg}%
  \BibitemOpen
  \bibfield  {author} {\bibinfo {author} {\bibfnamefont {R.}~\bibnamefont
  {Emparan}}\ and\ \bibinfo {author} {\bibfnamefont {H.~S.}\ \bibnamefont
  {Reall}},\ }\href {\doibase 10.12942/lrr-2008-6} {\bibfield  {journal}
  {\bibinfo  {journal} {Living Rev. Rel.}\ }\textbf {\bibinfo {volume} {11}},\
  \bibinfo {pages} {6} (\bibinfo {year} {2008})},\ \Eprint
  {http://arxiv.org/abs/0801.3471} {arXiv:0801.3471 [hep-th]} \BibitemShut
  {NoStop}%
\bibitem [{\citenamefont {Dubovsky}(2004)}]{Dubovsky:2004sg}%
  \BibitemOpen
  \bibfield  {author} {\bibinfo {author} {\bibfnamefont {S.~L.}\ \bibnamefont
  {Dubovsky}},\ }\href {\doibase 10.1088/1126-6708/2004/10/076} {\bibfield
  {journal} {\bibinfo  {journal} {JHEP}\ }\textbf {\bibinfo {volume} {10}},\
  \bibinfo {pages} {076} (\bibinfo {year} {2004})},\ \Eprint
  {http://arxiv.org/abs/hep-th/0409124} {arXiv:hep-th/0409124 [hep-th]}
  \BibitemShut {NoStop}%
\bibitem [{\citenamefont {Dubovsky}\ \emph {et~al.}(2006)\citenamefont
  {Dubovsky}, \citenamefont {Gregoire}, \citenamefont {Nicolis},\ and\
  \citenamefont {Rattazzi}}]{Dubovsky:2005xd}%
  \BibitemOpen
  \bibfield  {author} {\bibinfo {author} {\bibfnamefont {S.}~\bibnamefont
  {Dubovsky}}, \bibinfo {author} {\bibfnamefont {T.}~\bibnamefont {Gregoire}},
  \bibinfo {author} {\bibfnamefont {A.}~\bibnamefont {Nicolis}}, \ and\
  \bibinfo {author} {\bibfnamefont {R.}~\bibnamefont {Rattazzi}},\ }\href
  {\doibase 10.1088/1126-6708/2006/03/025} {\bibfield  {journal} {\bibinfo
  {journal} {JHEP}\ }\textbf {\bibinfo {volume} {03}},\ \bibinfo {pages} {025}
  (\bibinfo {year} {2006})},\ \Eprint {http://arxiv.org/abs/hep-th/0512260}
  {arXiv:hep-th/0512260 [hep-th]} \BibitemShut {NoStop}%
\bibitem [{\citenamefont {de~Rham}\ \emph {et~al.}(2011)\citenamefont
  {de~Rham}, \citenamefont {Gabadadze},\ and\ \citenamefont
  {Tolley}}]{deRham:2010kj}%
  \BibitemOpen
  \bibfield  {author} {\bibinfo {author} {\bibfnamefont {C.}~\bibnamefont
  {de~Rham}}, \bibinfo {author} {\bibfnamefont {G.}~\bibnamefont {Gabadadze}},
  \ and\ \bibinfo {author} {\bibfnamefont {A.~J.}\ \bibnamefont {Tolley}},\
  }\href {\doibase 10.1103/PhysRevLett.106.231101} {\bibfield  {journal}
  {\bibinfo  {journal} {Phys. Rev. Lett.}\ }\textbf {\bibinfo {volume} {106}},\
  \bibinfo {pages} {231101} (\bibinfo {year} {2011})},\ \Eprint
  {http://arxiv.org/abs/1011.1232} {arXiv:1011.1232 [hep-th]} \BibitemShut
  {NoStop}%
\bibitem [{\citenamefont {de~Rham}\ \emph {et~al.}(2012)\citenamefont
  {de~Rham}, \citenamefont {Gabadadze},\ and\ \citenamefont
  {Tolley}}]{deRham:2011rn}%
  \BibitemOpen
  \bibfield  {author} {\bibinfo {author} {\bibfnamefont {C.}~\bibnamefont
  {de~Rham}}, \bibinfo {author} {\bibfnamefont {G.}~\bibnamefont {Gabadadze}},
  \ and\ \bibinfo {author} {\bibfnamefont {A.~J.}\ \bibnamefont {Tolley}},\
  }\href {\doibase 10.1016/j.physletb.2012.03.081} {\bibfield  {journal}
  {\bibinfo  {journal} {Phys. Lett.}\ }\textbf {\bibinfo {volume} {B711}},\
  \bibinfo {pages} {190} (\bibinfo {year} {2012})},\ \Eprint
  {http://arxiv.org/abs/1107.3820} {arXiv:1107.3820 [hep-th]} \BibitemShut
  {NoStop}%
\bibitem [{\citenamefont {Ballesteros}\ \emph {et~al.}(2016)\citenamefont
  {Ballesteros}, \citenamefont {Comelli},\ and\ \citenamefont
  {Pilo}}]{Ballesteros:2016gwc}%
  \BibitemOpen
  \bibfield  {author} {\bibinfo {author} {\bibfnamefont {G.}~\bibnamefont
  {Ballesteros}}, \bibinfo {author} {\bibfnamefont {D.}~\bibnamefont
  {Comelli}}, \ and\ \bibinfo {author} {\bibfnamefont {L.}~\bibnamefont
  {Pilo}},\ }\href {\doibase 10.1103/PhysRevD.94.124023} {\bibfield  {journal}
  {\bibinfo  {journal} {Phys. Rev.}\ }\textbf {\bibinfo {volume} {D94}},\
  \bibinfo {pages} {124023} (\bibinfo {year} {2016})},\ \Eprint
  {http://arxiv.org/abs/1603.02956} {arXiv:1603.02956 [hep-th]} \BibitemShut
  {NoStop}%
\bibitem [{\citenamefont {Celoria}\ \emph {et~al.}(2017)\citenamefont
  {Celoria}, \citenamefont {Comelli},\ and\ \citenamefont
  {Pilo}}]{Celoria:2017bbh}%
  \BibitemOpen
  \bibfield  {author} {\bibinfo {author} {\bibfnamefont {M.}~\bibnamefont
  {Celoria}}, \bibinfo {author} {\bibfnamefont {D.}~\bibnamefont {Comelli}}, \
  and\ \bibinfo {author} {\bibfnamefont {L.}~\bibnamefont {Pilo}},\ }\href
  {\doibase 10.1088/1475-7516/2017/09/036} {\bibfield  {journal} {\bibinfo
  {journal} {JCAP}\ }\textbf {\bibinfo {volume} {1709}},\ \bibinfo {pages}
  {036} (\bibinfo {year} {2017})},\ \Eprint {http://arxiv.org/abs/1704.00322}
  {arXiv:1704.00322 [gr-qc]} \BibitemShut {NoStop}%
\bibitem [{\citenamefont {Kobayashi}\ \emph {et~al.}(2015)\citenamefont
  {Kobayashi}, \citenamefont {Watanabe},\ and\ \citenamefont
  {Yamauchi}}]{Kobayashi:2014ida}%
  \BibitemOpen
  \bibfield  {author} {\bibinfo {author} {\bibfnamefont {T.}~\bibnamefont
  {Kobayashi}}, \bibinfo {author} {\bibfnamefont {Y.}~\bibnamefont {Watanabe}},
  \ and\ \bibinfo {author} {\bibfnamefont {D.}~\bibnamefont {Yamauchi}},\
  }\href {\doibase 10.1103/PhysRevD.91.064013} {\bibfield  {journal} {\bibinfo
  {journal} {Phys. Rev.}\ }\textbf {\bibinfo {volume} {D91}},\ \bibinfo {pages}
  {064013} (\bibinfo {year} {2015})},\ \Eprint {http://arxiv.org/abs/1411.4130}
  {arXiv:1411.4130 [gr-qc]} \BibitemShut {NoStop}%
\bibitem [{\citenamefont {Koyama}\ and\ \citenamefont
  {Sakstein}(2015)}]{Koyama:2015oma}%
  \BibitemOpen
  \bibfield  {author} {\bibinfo {author} {\bibfnamefont {K.}~\bibnamefont
  {Koyama}}\ and\ \bibinfo {author} {\bibfnamefont {J.}~\bibnamefont
  {Sakstein}},\ }\href {\doibase 10.1103/PhysRevD.91.124066} {\bibfield
  {journal} {\bibinfo  {journal} {Phys. Rev.}\ }\textbf {\bibinfo {volume}
  {D91}},\ \bibinfo {pages} {124066} (\bibinfo {year} {2015})},\ \Eprint
  {http://arxiv.org/abs/1502.06872} {arXiv:1502.06872 [astro-ph.CO]}
  \BibitemShut {NoStop}%
\bibitem [{\citenamefont {Saito}\ \emph {et~al.}(2015)\citenamefont {Saito},
  \citenamefont {Yamauchi}, \citenamefont {Mizuno}, \citenamefont {Gleyzes},\
  and\ \citenamefont {Langlois}}]{Saito:2015fza}%
  \BibitemOpen
  \bibfield  {author} {\bibinfo {author} {\bibfnamefont {R.}~\bibnamefont
  {Saito}}, \bibinfo {author} {\bibfnamefont {D.}~\bibnamefont {Yamauchi}},
  \bibinfo {author} {\bibfnamefont {S.}~\bibnamefont {Mizuno}}, \bibinfo
  {author} {\bibfnamefont {J.}~\bibnamefont {Gleyzes}}, \ and\ \bibinfo
  {author} {\bibfnamefont {D.}~\bibnamefont {Langlois}},\ }\href {\doibase
  10.1088/1475-7516/2015/06/008} {\bibfield  {journal} {\bibinfo  {journal}
  {JCAP}\ }\textbf {\bibinfo {volume} {1506}},\ \bibinfo {pages} {008}
  (\bibinfo {year} {2015})},\ \Eprint {http://arxiv.org/abs/1503.01448}
  {arXiv:1503.01448 [gr-qc]} \BibitemShut {NoStop}%
\bibitem [{\citenamefont {Koyama}\ \emph
  {et~al.}(2011{\natexlab{a}})\citenamefont {Koyama}, \citenamefont {Niz},\
  and\ \citenamefont {Tasinato}}]{Koyama:2011xz}%
  \BibitemOpen
  \bibfield  {author} {\bibinfo {author} {\bibfnamefont {K.}~\bibnamefont
  {Koyama}}, \bibinfo {author} {\bibfnamefont {G.}~\bibnamefont {Niz}}, \ and\
  \bibinfo {author} {\bibfnamefont {G.}~\bibnamefont {Tasinato}},\ }\href
  {\doibase 10.1103/PhysRevLett.107.131101} {\bibfield  {journal} {\bibinfo
  {journal} {Phys. Rev. Lett.}\ }\textbf {\bibinfo {volume} {107}},\ \bibinfo
  {pages} {131101} (\bibinfo {year} {2011}{\natexlab{a}})},\ \Eprint
  {http://arxiv.org/abs/1103.4708} {arXiv:1103.4708 [hep-th]} \BibitemShut
  {NoStop}%
\bibitem [{\citenamefont {Koyama}\ \emph
  {et~al.}(2011{\natexlab{b}})\citenamefont {Koyama}, \citenamefont {Niz},\
  and\ \citenamefont {Tasinato}}]{Koyama:2011yg}%
  \BibitemOpen
  \bibfield  {author} {\bibinfo {author} {\bibfnamefont {K.}~\bibnamefont
  {Koyama}}, \bibinfo {author} {\bibfnamefont {G.}~\bibnamefont {Niz}}, \ and\
  \bibinfo {author} {\bibfnamefont {G.}~\bibnamefont {Tasinato}},\ }\href
  {\doibase 10.1103/PhysRevD.84.064033} {\bibfield  {journal} {\bibinfo
  {journal} {Phys. Rev.}\ }\textbf {\bibinfo {volume} {D84}},\ \bibinfo {pages}
  {064033} (\bibinfo {year} {2011}{\natexlab{b}})},\ \Eprint
  {http://arxiv.org/abs/1104.2143} {arXiv:1104.2143 [hep-th]} \BibitemShut
  {NoStop}%
\bibitem [{\citenamefont {Comelli}\ \emph {et~al.}(2012)\citenamefont
  {Comelli}, \citenamefont {Crisostomi}, \citenamefont {Nesti},\ and\
  \citenamefont {Pilo}}]{Comelli:2011wq}%
  \BibitemOpen
  \bibfield  {author} {\bibinfo {author} {\bibfnamefont {D.}~\bibnamefont
  {Comelli}}, \bibinfo {author} {\bibfnamefont {M.}~\bibnamefont {Crisostomi}},
  \bibinfo {author} {\bibfnamefont {F.}~\bibnamefont {Nesti}}, \ and\ \bibinfo
  {author} {\bibfnamefont {L.}~\bibnamefont {Pilo}},\ }\href {\doibase
  10.1103/PhysRevD.85.024044} {\bibfield  {journal} {\bibinfo  {journal} {Phys.
  Rev.}\ }\textbf {\bibinfo {volume} {D85}},\ \bibinfo {pages} {024044}
  (\bibinfo {year} {2012})},\ \Eprint {http://arxiv.org/abs/1110.4967}
  {arXiv:1110.4967 [hep-th]} \BibitemShut {NoStop}%
\bibitem [{\citenamefont {Berezhiani}\ \emph {et~al.}(2012)\citenamefont
  {Berezhiani}, \citenamefont {Chkareuli}, \citenamefont {de~Rham},
  \citenamefont {Gabadadze},\ and\ \citenamefont {Tolley}}]{Berezhiani:2011mt}%
  \BibitemOpen
  \bibfield  {author} {\bibinfo {author} {\bibfnamefont {L.}~\bibnamefont
  {Berezhiani}}, \bibinfo {author} {\bibfnamefont {G.}~\bibnamefont
  {Chkareuli}}, \bibinfo {author} {\bibfnamefont {C.}~\bibnamefont {de~Rham}},
  \bibinfo {author} {\bibfnamefont {G.}~\bibnamefont {Gabadadze}}, \ and\
  \bibinfo {author} {\bibfnamefont {A.~J.}\ \bibnamefont {Tolley}},\ }\href
  {\doibase 10.1103/PhysRevD.85.044024} {\bibfield  {journal} {\bibinfo
  {journal} {Phys. Rev.}\ }\textbf {\bibinfo {volume} {D85}},\ \bibinfo {pages}
  {044024} (\bibinfo {year} {2012})},\ \Eprint {http://arxiv.org/abs/1111.3613}
  {arXiv:1111.3613 [hep-th]} \BibitemShut {NoStop}%
\bibitem [{\citenamefont {Volkov}(2013)}]{Volkov:2013roa}%
  \BibitemOpen
  \bibfield  {author} {\bibinfo {author} {\bibfnamefont {M.~S.}\ \bibnamefont
  {Volkov}},\ }\href {\doibase 10.1088/0264-9381/30/18/184009} {\bibfield
  {journal} {\bibinfo  {journal} {Class. Quant. Grav.}\ }\textbf {\bibinfo
  {volume} {30}},\ \bibinfo {pages} {184009} (\bibinfo {year} {2013})},\
  \Eprint {http://arxiv.org/abs/1304.0238} {arXiv:1304.0238 [hep-th]}
  \BibitemShut {NoStop}%
\bibitem [{\citenamefont {Jackiw}\ and\ \citenamefont
  {Pi}(2003)}]{Jackiw:2003pm}%
  \BibitemOpen
  \bibfield  {author} {\bibinfo {author} {\bibfnamefont {R.}~\bibnamefont
  {Jackiw}}\ and\ \bibinfo {author} {\bibfnamefont {S.~Y.}\ \bibnamefont
  {Pi}},\ }\href {\doibase 10.1103/PhysRevD.68.104012} {\bibfield  {journal}
  {\bibinfo  {journal} {Phys. Rev.}\ }\textbf {\bibinfo {volume} {D68}},\
  \bibinfo {pages} {104012} (\bibinfo {year} {2003})},\ \Eprint
  {http://arxiv.org/abs/gr-qc/0308071} {arXiv:gr-qc/0308071 [gr-qc]}
  \BibitemShut {NoStop}%
\bibitem [{\citenamefont {Smith}\ \emph {et~al.}(2008)\citenamefont {Smith},
  \citenamefont {Erickcek}, \citenamefont {Caldwell},\ and\ \citenamefont
  {Kamionkowski}}]{Smith:2007jm}%
  \BibitemOpen
  \bibfield  {author} {\bibinfo {author} {\bibfnamefont {T.~L.}\ \bibnamefont
  {Smith}}, \bibinfo {author} {\bibfnamefont {A.~L.}\ \bibnamefont {Erickcek}},
  \bibinfo {author} {\bibfnamefont {R.~R.}\ \bibnamefont {Caldwell}}, \ and\
  \bibinfo {author} {\bibfnamefont {M.}~\bibnamefont {Kamionkowski}},\ }\href
  {\doibase 10.1103/PhysRevD.77.024015} {\bibfield  {journal} {\bibinfo
  {journal} {Phys. Rev.}\ }\textbf {\bibinfo {volume} {D77}},\ \bibinfo {pages}
  {024015} (\bibinfo {year} {2008})},\ \Eprint {http://arxiv.org/abs/0708.0001}
  {arXiv:0708.0001 [astro-ph]} \BibitemShut {NoStop}%
\bibitem [{\citenamefont {Grumiller}\ and\ \citenamefont
  {Yunes}(2008)}]{Grumiller:2007rv}%
  \BibitemOpen
  \bibfield  {author} {\bibinfo {author} {\bibfnamefont {D.}~\bibnamefont
  {Grumiller}}\ and\ \bibinfo {author} {\bibfnamefont {N.}~\bibnamefont
  {Yunes}},\ }\href {\doibase 10.1103/PhysRevD.77.044015} {\bibfield  {journal}
  {\bibinfo  {journal} {Phys. Rev.}\ }\textbf {\bibinfo {volume} {D77}},\
  \bibinfo {pages} {044015} (\bibinfo {year} {2008})},\ \Eprint
  {http://arxiv.org/abs/0711.1868} {arXiv:0711.1868 [gr-qc]} \BibitemShut
  {NoStop}%
\bibitem [{\citenamefont {Yunes}\ and\ \citenamefont
  {Pretorius}(2009)}]{Yunes:2009hc}%
  \BibitemOpen
  \bibfield  {author} {\bibinfo {author} {\bibfnamefont {N.}~\bibnamefont
  {Yunes}}\ and\ \bibinfo {author} {\bibfnamefont {F.}~\bibnamefont
  {Pretorius}},\ }\href {\doibase 10.1103/PhysRevD.79.084043} {\bibfield
  {journal} {\bibinfo  {journal} {Phys. Rev.}\ }\textbf {\bibinfo {volume}
  {D79}},\ \bibinfo {pages} {084043} (\bibinfo {year} {2009})},\ \Eprint
  {http://arxiv.org/abs/0902.4669} {arXiv:0902.4669 [gr-qc]} \BibitemShut
  {NoStop}%
\bibitem [{\citenamefont {Konno}\ \emph {et~al.}(2009)\citenamefont {Konno},
  \citenamefont {Matsuyama},\ and\ \citenamefont {Tanda}}]{Konno:2009kg}%
  \BibitemOpen
  \bibfield  {author} {\bibinfo {author} {\bibfnamefont {K.}~\bibnamefont
  {Konno}}, \bibinfo {author} {\bibfnamefont {T.}~\bibnamefont {Matsuyama}}, \
  and\ \bibinfo {author} {\bibfnamefont {S.}~\bibnamefont {Tanda}},\ }\href
  {\doibase 10.1143/PTP.122.561} {\bibfield  {journal} {\bibinfo  {journal}
  {Prog. Theor. Phys.}\ }\textbf {\bibinfo {volume} {122}},\ \bibinfo {pages}
  {561} (\bibinfo {year} {2009})},\ \Eprint {http://arxiv.org/abs/0902.4767}
  {arXiv:0902.4767 [gr-qc]} \BibitemShut {NoStop}%
\bibitem [{\citenamefont {Motohashi}\ and\ \citenamefont
  {Suyama}(2011)}]{Motohashi:2011pw}%
  \BibitemOpen
  \bibfield  {author} {\bibinfo {author} {\bibfnamefont {H.}~\bibnamefont
  {Motohashi}}\ and\ \bibinfo {author} {\bibfnamefont {T.}~\bibnamefont
  {Suyama}},\ }\href {\doibase 10.1103/PhysRevD.84.084041} {\bibfield
  {journal} {\bibinfo  {journal} {Phys. Rev.}\ }\textbf {\bibinfo {volume}
  {D84}},\ \bibinfo {pages} {084041} (\bibinfo {year} {2011})},\ \Eprint
  {http://arxiv.org/abs/1107.3705} {arXiv:1107.3705 [gr-qc]} \BibitemShut
  {NoStop}%
\bibitem [{\citenamefont {Motohashi}\ and\ \citenamefont
  {Suyama}(2012)}]{Motohashi:2011ds}%
  \BibitemOpen
  \bibfield  {author} {\bibinfo {author} {\bibfnamefont {H.}~\bibnamefont
  {Motohashi}}\ and\ \bibinfo {author} {\bibfnamefont {T.}~\bibnamefont
  {Suyama}},\ }\href {\doibase 10.1103/PhysRevD.85.044054} {\bibfield
  {journal} {\bibinfo  {journal} {Phys. Rev.}\ }\textbf {\bibinfo {volume}
  {D85}},\ \bibinfo {pages} {044054} (\bibinfo {year} {2012})},\ \Eprint
  {http://arxiv.org/abs/1110.6241} {arXiv:1110.6241 [gr-qc]} \BibitemShut
  {NoStop}%
\bibitem [{\citenamefont {Ayzenberg}\ \emph {et~al.}(2014)\citenamefont
  {Ayzenberg}, \citenamefont {Yagi},\ and\ \citenamefont
  {Yunes}}]{Ayzenberg:2013wua}%
  \BibitemOpen
  \bibfield  {author} {\bibinfo {author} {\bibfnamefont {D.}~\bibnamefont
  {Ayzenberg}}, \bibinfo {author} {\bibfnamefont {K.}~\bibnamefont {Yagi}}, \
  and\ \bibinfo {author} {\bibfnamefont {N.}~\bibnamefont {Yunes}},\ }\href
  {\doibase 10.1103/PhysRevD.89.044023} {\bibfield  {journal} {\bibinfo
  {journal} {Phys. Rev.}\ }\textbf {\bibinfo {volume} {D89}},\ \bibinfo {pages}
  {044023} (\bibinfo {year} {2014})},\ \Eprint {http://arxiv.org/abs/1310.6392}
  {arXiv:1310.6392 [gr-qc]} \BibitemShut {NoStop}%
\bibitem [{\citenamefont {Konno}\ and\ \citenamefont
  {Takahashi}(2014)}]{Konno:2014qua}%
  \BibitemOpen
  \bibfield  {author} {\bibinfo {author} {\bibfnamefont {K.}~\bibnamefont
  {Konno}}\ and\ \bibinfo {author} {\bibfnamefont {R.}~\bibnamefont
  {Takahashi}},\ }\href {\doibase 10.1103/PhysRevD.90.064011} {\bibfield
  {journal} {\bibinfo  {journal} {Phys. Rev.}\ }\textbf {\bibinfo {volume}
  {D90}},\ \bibinfo {pages} {064011} (\bibinfo {year} {2014})},\ \Eprint
  {http://arxiv.org/abs/1406.0957} {arXiv:1406.0957 [gr-qc]} \BibitemShut
  {NoStop}%
\bibitem [{\citenamefont {Stein}(2014)}]{Stein:2014xba}%
  \BibitemOpen
  \bibfield  {author} {\bibinfo {author} {\bibfnamefont {L.~C.}\ \bibnamefont
  {Stein}},\ }\href {\doibase 10.1103/PhysRevD.90.044061} {\bibfield  {journal}
  {\bibinfo  {journal} {Phys. Rev.}\ }\textbf {\bibinfo {volume} {D90}},\
  \bibinfo {pages} {044061} (\bibinfo {year} {2014})},\ \Eprint
  {http://arxiv.org/abs/1407.2350} {arXiv:1407.2350 [gr-qc]} \BibitemShut
  {NoStop}%
\bibitem [{\citenamefont {Deffayet}\ \emph {et~al.}(2010)\citenamefont
  {Deffayet}, \citenamefont {Deser},\ and\ \citenamefont
  {Esposito-Farese}}]{Deffayet:2010zh}%
  \BibitemOpen
  \bibfield  {author} {\bibinfo {author} {\bibfnamefont {C.}~\bibnamefont
  {Deffayet}}, \bibinfo {author} {\bibfnamefont {S.}~\bibnamefont {Deser}}, \
  and\ \bibinfo {author} {\bibfnamefont {G.}~\bibnamefont {Esposito-Farese}},\
  }\href {\doibase 10.1103/PhysRevD.82.061501} {\bibfield  {journal} {\bibinfo
  {journal} {Phys. Rev.}\ }\textbf {\bibinfo {volume} {D82}},\ \bibinfo {pages}
  {061501} (\bibinfo {year} {2010})},\ \Eprint {http://arxiv.org/abs/1007.5278}
  {arXiv:1007.5278 [gr-qc]} \BibitemShut {NoStop}%
\bibitem [{\citenamefont {Padilla}\ \emph {et~al.}(2010)\citenamefont
  {Padilla}, \citenamefont {Saffin},\ and\ \citenamefont
  {Zhou}}]{Padilla:2010de}%
  \BibitemOpen
  \bibfield  {author} {\bibinfo {author} {\bibfnamefont {A.}~\bibnamefont
  {Padilla}}, \bibinfo {author} {\bibfnamefont {P.~M.}\ \bibnamefont {Saffin}},
  \ and\ \bibinfo {author} {\bibfnamefont {S.-Y.}\ \bibnamefont {Zhou}},\
  }\href {\doibase 10.1007/JHEP12(2010)031} {\bibfield  {journal} {\bibinfo
  {journal} {JHEP}\ }\textbf {\bibinfo {volume} {12}},\ \bibinfo {pages} {031}
  (\bibinfo {year} {2010})},\ \Eprint {http://arxiv.org/abs/1007.5424}
  {arXiv:1007.5424 [hep-th]} \BibitemShut {NoStop}%
\bibitem [{\citenamefont {Padilla}\ \emph
  {et~al.}(2011{\natexlab{a}})\citenamefont {Padilla}, \citenamefont {Saffin},\
  and\ \citenamefont {Zhou}}]{Padilla:2010ir}%
  \BibitemOpen
  \bibfield  {author} {\bibinfo {author} {\bibfnamefont {A.}~\bibnamefont
  {Padilla}}, \bibinfo {author} {\bibfnamefont {P.~M.}\ \bibnamefont {Saffin}},
  \ and\ \bibinfo {author} {\bibfnamefont {S.-Y.}\ \bibnamefont {Zhou}},\
  }\href {\doibase 10.1103/PhysRevD.83.045009} {\bibfield  {journal} {\bibinfo
  {journal} {Phys. Rev.}\ }\textbf {\bibinfo {volume} {D83}},\ \bibinfo {pages}
  {045009} (\bibinfo {year} {2011}{\natexlab{a}})},\ \Eprint
  {http://arxiv.org/abs/1008.0745} {arXiv:1008.0745 [hep-th]} \BibitemShut
  {NoStop}%
\bibitem [{\citenamefont {Padilla}\ \emph
  {et~al.}(2011{\natexlab{b}})\citenamefont {Padilla}, \citenamefont {Saffin},\
  and\ \citenamefont {Zhou}}]{Padilla:2010tj}%
  \BibitemOpen
  \bibfield  {author} {\bibinfo {author} {\bibfnamefont {A.}~\bibnamefont
  {Padilla}}, \bibinfo {author} {\bibfnamefont {P.~M.}\ \bibnamefont {Saffin}},
  \ and\ \bibinfo {author} {\bibfnamefont {S.-Y.}\ \bibnamefont {Zhou}},\
  }\href {\doibase 10.1007/JHEP01(2011)099} {\bibfield  {journal} {\bibinfo
  {journal} {JHEP}\ }\textbf {\bibinfo {volume} {01}},\ \bibinfo {pages} {099}
  (\bibinfo {year} {2011}{\natexlab{b}})},\ \Eprint
  {http://arxiv.org/abs/1008.3312} {arXiv:1008.3312 [hep-th]} \BibitemShut
  {NoStop}%
\bibitem [{\citenamefont {Trodden}\ and\ \citenamefont
  {Hinterbichler}(2011)}]{Trodden:2011xh}%
  \BibitemOpen
  \bibfield  {author} {\bibinfo {author} {\bibfnamefont {M.}~\bibnamefont
  {Trodden}}\ and\ \bibinfo {author} {\bibfnamefont {K.}~\bibnamefont
  {Hinterbichler}},\ }\href {\doibase 10.1088/0264-9381/28/20/204003}
  {\bibfield  {journal} {\bibinfo  {journal} {Class. Quant. Grav.}\ }\textbf
  {\bibinfo {volume} {28}},\ \bibinfo {pages} {204003} (\bibinfo {year}
  {2011})},\ \Eprint {http://arxiv.org/abs/1104.2088} {arXiv:1104.2088
  [hep-th]} \BibitemShut {NoStop}%
\bibitem [{\citenamefont {Padilla}\ and\ \citenamefont
  {Sivanesan}(2013)}]{Padilla:2012dx}%
  \BibitemOpen
  \bibfield  {author} {\bibinfo {author} {\bibfnamefont {A.}~\bibnamefont
  {Padilla}}\ and\ \bibinfo {author} {\bibfnamefont {V.}~\bibnamefont
  {Sivanesan}},\ }\href {\doibase 10.1007/JHEP04(2013)032} {\bibfield
  {journal} {\bibinfo  {journal} {JHEP}\ }\textbf {\bibinfo {volume} {04}},\
  \bibinfo {pages} {032} (\bibinfo {year} {2013})},\ \Eprint
  {http://arxiv.org/abs/1210.4026} {arXiv:1210.4026 [gr-qc]} \BibitemShut
  {NoStop}%
\bibitem [{\citenamefont {Kobayashi}\ \emph {et~al.}(2013)\citenamefont
  {Kobayashi}, \citenamefont {Tanahashi},\ and\ \citenamefont
  {Yamaguchi}}]{Kobayashi:2013ina}%
  \BibitemOpen
  \bibfield  {author} {\bibinfo {author} {\bibfnamefont {T.}~\bibnamefont
  {Kobayashi}}, \bibinfo {author} {\bibfnamefont {N.}~\bibnamefont
  {Tanahashi}}, \ and\ \bibinfo {author} {\bibfnamefont {M.}~\bibnamefont
  {Yamaguchi}},\ }\href {\doibase 10.1103/PhysRevD.88.083504} {\bibfield
  {journal} {\bibinfo  {journal} {Phys. Rev.}\ }\textbf {\bibinfo {volume}
  {D88}},\ \bibinfo {pages} {083504} (\bibinfo {year} {2013})},\ \Eprint
  {http://arxiv.org/abs/1308.4798} {arXiv:1308.4798 [hep-th]} \BibitemShut
  {NoStop}%
\bibitem [{\citenamefont {Ohashi}\ \emph {et~al.}(2015)\citenamefont {Ohashi},
  \citenamefont {Tanahashi}, \citenamefont {Kobayashi},\ and\ \citenamefont
  {Yamaguchi}}]{Ohashi:2015fma}%
  \BibitemOpen
  \bibfield  {author} {\bibinfo {author} {\bibfnamefont {S.}~\bibnamefont
  {Ohashi}}, \bibinfo {author} {\bibfnamefont {N.}~\bibnamefont {Tanahashi}},
  \bibinfo {author} {\bibfnamefont {T.}~\bibnamefont {Kobayashi}}, \ and\
  \bibinfo {author} {\bibfnamefont {M.}~\bibnamefont {Yamaguchi}},\ }\href
  {\doibase 10.1007/JHEP07(2015)008} {\bibfield  {journal} {\bibinfo  {journal}
  {JHEP}\ }\textbf {\bibinfo {volume} {07}},\ \bibinfo {pages} {008} (\bibinfo
  {year} {2015})},\ \Eprint {http://arxiv.org/abs/1505.06029} {arXiv:1505.06029
  [gr-qc]} \BibitemShut {NoStop}%
\bibitem [{\citenamefont {Allys}(2017)}]{Allys:2016hfl}%
  \BibitemOpen
  \bibfield  {author} {\bibinfo {author} {\bibfnamefont {E.}~\bibnamefont
  {Allys}},\ }\href {\doibase 10.1103/PhysRevD.95.064051} {\bibfield  {journal}
  {\bibinfo  {journal} {Phys. Rev.}\ }\textbf {\bibinfo {volume} {D95}},\
  \bibinfo {pages} {064051} (\bibinfo {year} {2017})},\ \Eprint
  {http://arxiv.org/abs/1612.01972} {arXiv:1612.01972 [hep-th]} \BibitemShut
  {NoStop}%
\bibitem [{\citenamefont {Afshordi}\ \emph {et~al.}(2007)\citenamefont
  {Afshordi}, \citenamefont {Chung},\ and\ \citenamefont
  {Geshnizjani}}]{Afshordi:2006ad}%
  \BibitemOpen
  \bibfield  {author} {\bibinfo {author} {\bibfnamefont {N.}~\bibnamefont
  {Afshordi}}, \bibinfo {author} {\bibfnamefont {D.~J.~H.}\ \bibnamefont
  {Chung}}, \ and\ \bibinfo {author} {\bibfnamefont {G.}~\bibnamefont
  {Geshnizjani}},\ }\href {\doibase 10.1103/PhysRevD.75.083513} {\bibfield
  {journal} {\bibinfo  {journal} {Phys. Rev.}\ }\textbf {\bibinfo {volume}
  {D75}},\ \bibinfo {pages} {083513} (\bibinfo {year} {2007})},\ \Eprint
  {http://arxiv.org/abs/hep-th/0609150} {arXiv:hep-th/0609150 [hep-th]}
  \BibitemShut {NoStop}%
\bibitem [{\citenamefont {de~Rham}\ and\ \citenamefont
  {Motohashi}(2017)}]{deRham:2016ged}%
  \BibitemOpen
  \bibfield  {author} {\bibinfo {author} {\bibfnamefont {C.}~\bibnamefont
  {de~Rham}}\ and\ \bibinfo {author} {\bibfnamefont {H.}~\bibnamefont
  {Motohashi}},\ }\href {\doibase 10.1103/PhysRevD.95.064008} {\bibfield
  {journal} {\bibinfo  {journal} {Phys. Rev.}\ }\textbf {\bibinfo {volume}
  {D95}},\ \bibinfo {pages} {064008} (\bibinfo {year} {2017})},\ \Eprint
  {http://arxiv.org/abs/1611.05038} {arXiv:1611.05038 [hep-th]} \BibitemShut
  {NoStop}%
\bibitem [{\citenamefont {Doneva}\ and\ \citenamefont
  {Yazadjiev}(2018)}]{Doneva:2017bvd}%
  \BibitemOpen
  \bibfield  {author} {\bibinfo {author} {\bibfnamefont {D.~D.}\ \bibnamefont
  {Doneva}}\ and\ \bibinfo {author} {\bibfnamefont {S.~S.}\ \bibnamefont
  {Yazadjiev}},\ }\href {\doibase 10.1103/PhysRevLett.120.131103} {\bibfield
  {journal} {\bibinfo  {journal} {Phys. Rev. Lett.}\ }\textbf {\bibinfo
  {volume} {120}},\ \bibinfo {pages} {131103} (\bibinfo {year} {2018})},\
  \Eprint {http://arxiv.org/abs/1711.01187} {arXiv:1711.01187 [gr-qc]}
  \BibitemShut {NoStop}%
\bibitem [{\citenamefont {Silva}\ \emph {et~al.}(2018)\citenamefont {Silva},
  \citenamefont {Sakstein}, \citenamefont {Gualtieri}, \citenamefont
  {Sotiriou},\ and\ \citenamefont {Berti}}]{Silva:2017uqg}%
  \BibitemOpen
  \bibfield  {author} {\bibinfo {author} {\bibfnamefont {H.~O.}\ \bibnamefont
  {Silva}}, \bibinfo {author} {\bibfnamefont {J.}~\bibnamefont {Sakstein}},
  \bibinfo {author} {\bibfnamefont {L.}~\bibnamefont {Gualtieri}}, \bibinfo
  {author} {\bibfnamefont {T.~P.}\ \bibnamefont {Sotiriou}}, \ and\ \bibinfo
  {author} {\bibfnamefont {E.}~\bibnamefont {Berti}},\ }\href {\doibase
  10.1103/PhysRevLett.120.131104} {\bibfield  {journal} {\bibinfo  {journal}
  {Phys. Rev. Lett.}\ }\textbf {\bibinfo {volume} {120}},\ \bibinfo {pages}
  {131104} (\bibinfo {year} {2018})},\ \Eprint
  {http://arxiv.org/abs/1711.02080} {arXiv:1711.02080 [gr-qc]} \BibitemShut
  {NoStop}%
\bibitem [{\citenamefont {Antoniou}\ \emph
  {et~al.}(2018{\natexlab{a}})\citenamefont {Antoniou}, \citenamefont
  {Bakopoulos},\ and\ \citenamefont {Kanti}}]{Antoniou:2017acq}%
  \BibitemOpen
  \bibfield  {author} {\bibinfo {author} {\bibfnamefont {G.}~\bibnamefont
  {Antoniou}}, \bibinfo {author} {\bibfnamefont {A.}~\bibnamefont
  {Bakopoulos}}, \ and\ \bibinfo {author} {\bibfnamefont {P.}~\bibnamefont
  {Kanti}},\ }\href {\doibase 10.1103/PhysRevLett.120.131102} {\bibfield
  {journal} {\bibinfo  {journal} {Phys. Rev. Lett.}\ }\textbf {\bibinfo
  {volume} {120}},\ \bibinfo {pages} {131102} (\bibinfo {year}
  {2018}{\natexlab{a}})},\ \Eprint {http://arxiv.org/abs/1711.03390}
  {arXiv:1711.03390 [hep-th]} \BibitemShut {NoStop}%
\bibitem [{\citenamefont {Antoniou}\ \emph
  {et~al.}(2018{\natexlab{b}})\citenamefont {Antoniou}, \citenamefont
  {Bakopoulos},\ and\ \citenamefont {Kanti}}]{Antoniou:2017hxj}%
  \BibitemOpen
  \bibfield  {author} {\bibinfo {author} {\bibfnamefont {G.}~\bibnamefont
  {Antoniou}}, \bibinfo {author} {\bibfnamefont {A.}~\bibnamefont
  {Bakopoulos}}, \ and\ \bibinfo {author} {\bibfnamefont {P.}~\bibnamefont
  {Kanti}},\ }\href {\doibase 10.1103/PhysRevD.97.084037} {\bibfield  {journal}
  {\bibinfo  {journal} {Phys. Rev.}\ }\textbf {\bibinfo {volume} {D97}},\
  \bibinfo {pages} {084037} (\bibinfo {year} {2018}{\natexlab{b}})},\ \Eprint
  {http://arxiv.org/abs/1711.07431} {arXiv:1711.07431 [hep-th]} \BibitemShut
  {NoStop}%
\bibitem [{\citenamefont {Kobayashi}\ \emph {et~al.}(2012)\citenamefont
  {Kobayashi}, \citenamefont {Motohashi},\ and\ \citenamefont
  {Suyama}}]{Kobayashi:2012kh}%
  \BibitemOpen
  \bibfield  {author} {\bibinfo {author} {\bibfnamefont {T.}~\bibnamefont
  {Kobayashi}}, \bibinfo {author} {\bibfnamefont {H.}~\bibnamefont
  {Motohashi}}, \ and\ \bibinfo {author} {\bibfnamefont {T.}~\bibnamefont
  {Suyama}},\ }\href {\doibase 10.1103/PhysRevD.96.109903,
  10.1103/PhysRevD.85.084025} {\bibfield  {journal} {\bibinfo  {journal} {Phys.
  Rev.}\ }\textbf {\bibinfo {volume} {D85}},\ \bibinfo {pages} {084025}
  (\bibinfo {year} {2012})},\ \bibinfo {note} {[Erratum: Phys.
  Rev.D96,no.10,109903(2017)]},\ \Eprint {http://arxiv.org/abs/1202.4893}
  {arXiv:1202.4893 [gr-qc]} \BibitemShut {NoStop}%
\bibitem [{\citenamefont {Kobayashi}\ \emph {et~al.}(2014)\citenamefont
  {Kobayashi}, \citenamefont {Motohashi},\ and\ \citenamefont
  {Suyama}}]{Kobayashi:2014wsa}%
  \BibitemOpen
  \bibfield  {author} {\bibinfo {author} {\bibfnamefont {T.}~\bibnamefont
  {Kobayashi}}, \bibinfo {author} {\bibfnamefont {H.}~\bibnamefont
  {Motohashi}}, \ and\ \bibinfo {author} {\bibfnamefont {T.}~\bibnamefont
  {Suyama}},\ }\href {\doibase 10.1103/PhysRevD.89.084042} {\bibfield
  {journal} {\bibinfo  {journal} {Phys. Rev.}\ }\textbf {\bibinfo {volume}
  {D89}},\ \bibinfo {pages} {084042} (\bibinfo {year} {2014})},\ \Eprint
  {http://arxiv.org/abs/1402.6740} {arXiv:1402.6740 [gr-qc]} \BibitemShut
  {NoStop}%
\bibitem [{\citenamefont {Motohashi}\ \emph
  {et~al.}(2016{\natexlab{b}})\citenamefont {Motohashi}, \citenamefont
  {Suyama},\ and\ \citenamefont {Takahashi}}]{Motohashi:2016prk}%
  \BibitemOpen
  \bibfield  {author} {\bibinfo {author} {\bibfnamefont {H.}~\bibnamefont
  {Motohashi}}, \bibinfo {author} {\bibfnamefont {T.}~\bibnamefont {Suyama}}, \
  and\ \bibinfo {author} {\bibfnamefont {K.}~\bibnamefont {Takahashi}},\ }\href
  {\doibase 10.1103/PhysRevD.94.124021} {\bibfield  {journal} {\bibinfo
  {journal} {Phys. Rev.}\ }\textbf {\bibinfo {volume} {D94}},\ \bibinfo {pages}
  {124021} (\bibinfo {year} {2016}{\natexlab{b}})},\ \Eprint
  {http://arxiv.org/abs/1608.00071} {arXiv:1608.00071 [gr-qc]} \BibitemShut
  {NoStop}%
\bibitem [{\citenamefont {Lehebel}\ \emph {et~al.}(2017)\citenamefont
  {Lehebel}, \citenamefont {Babichev},\ and\ \citenamefont
  {Charmousis}}]{Lehebel:2017fag}%
  \BibitemOpen
  \bibfield  {author} {\bibinfo {author} {\bibfnamefont {A.}~\bibnamefont
  {Lehebel}}, \bibinfo {author} {\bibfnamefont {E.}~\bibnamefont {Babichev}}, \
  and\ \bibinfo {author} {\bibfnamefont {C.}~\bibnamefont {Charmousis}},\
  }\href {\doibase 10.1088/1475-7516/2017/07/037} {\bibfield  {journal}
  {\bibinfo  {journal} {JCAP}\ }\textbf {\bibinfo {volume} {1707}},\ \bibinfo
  {pages} {037} (\bibinfo {year} {2017})},\ \Eprint
  {http://arxiv.org/abs/1706.04989} {arXiv:1706.04989 [gr-qc]} \BibitemShut
  {NoStop}%
\bibitem [{\citenamefont {Damour}\ and\ \citenamefont
  {Esposito-Farese}(1993)}]{Damour:1993hw}%
  \BibitemOpen
  \bibfield  {author} {\bibinfo {author} {\bibfnamefont {T.}~\bibnamefont
  {Damour}}\ and\ \bibinfo {author} {\bibfnamefont {G.}~\bibnamefont
  {Esposito-Farese}},\ }\href {\doibase 10.1103/PhysRevLett.70.2220} {\bibfield
   {journal} {\bibinfo  {journal} {Phys. Rev. Lett.}\ }\textbf {\bibinfo
  {volume} {70}},\ \bibinfo {pages} {2220} (\bibinfo {year}
  {1993})}\BibitemShut {NoStop}%
\bibitem [{\citenamefont {Damour}\ and\ \citenamefont
  {Esposito-Farese}(1996)}]{Damour:1996ke}%
  \BibitemOpen
  \bibfield  {author} {\bibinfo {author} {\bibfnamefont {T.}~\bibnamefont
  {Damour}}\ and\ \bibinfo {author} {\bibfnamefont {G.}~\bibnamefont
  {Esposito-Farese}},\ }\href {\doibase 10.1103/PhysRevD.54.1474} {\bibfield
  {journal} {\bibinfo  {journal} {Phys. Rev.}\ }\textbf {\bibinfo {volume}
  {D54}},\ \bibinfo {pages} {1474} (\bibinfo {year} {1996})},\ \Eprint
  {http://arxiv.org/abs/gr-qc/9602056} {arXiv:gr-qc/9602056 [gr-qc]}
  \BibitemShut {NoStop}%
\bibitem [{\citenamefont {Kobayashi}\ and\ \citenamefont
  {Tanahashi}(2014)}]{Kobayashi:2014eva}%
  \BibitemOpen
  \bibfield  {author} {\bibinfo {author} {\bibfnamefont {T.}~\bibnamefont
  {Kobayashi}}\ and\ \bibinfo {author} {\bibfnamefont {N.}~\bibnamefont
  {Tanahashi}},\ }\href {\doibase 10.1093/ptep/ptu096} {\bibfield  {journal}
  {\bibinfo  {journal} {PTEP}\ }\textbf {\bibinfo {volume} {2014}},\ \bibinfo
  {pages} {073E02} (\bibinfo {year} {2014})},\ \Eprint
  {http://arxiv.org/abs/1403.4364} {arXiv:1403.4364 [gr-qc]} \BibitemShut
  {NoStop}%
\end{thebibliography}%

\end{document}